\documentclass[12pt]{article}
\pdfoutput=1
\usepackage{amsmath}
\usepackage{amssymb}
\usepackage{graphicx}
\usepackage[T1]{fontenc}
\usepackage{authblk}
\usepackage{subfigure}
\usepackage{color} 
\usepackage[hang, small, bf, margin=20pt, tableposition=top]{caption} 
\usepackage[square, comma, numbers, sort&compress]{natbib}
\usepackage[labelfont=bf,labelsep=period,justification=raggedright]{caption}
\title{Reconstructing transmission trees for communicable diseases using densely sampled genetic data}
\author[1,2 *]{\small Colin J. Worby} 
\author[1]{Philip D. O'Neill}
\author[1]{Theodore Kypraios}
\author[3]{Julie V. Robotham}
\author[4]{Daniela De~Angelis}
\author[5]{Edward J. P. Cartwright}
\author[5,6]{Sharon J. Peacock}
\author[7,8]{Ben S. Cooper}
\affil[1]{\footnotesize School of Mathematical Sciences, University of Nottingham, Nottingham, UK}
\affil[2]{Center for Communicable Disease Dynamics, Harvard TH Chan School of Public Health, Boston, USA}
\affil[3]{Modelling \& Economics Unit, Public Health England, London, UK}
\affil[4]{MRC Biostatistics Unit, University of Cambridge, Cambridge, UK}
\affil[5]{Department of Medicine, University of Cambridge, Addenbrooke's Hospital, Cambridge, UK}
\affil[6]{Wellcome Trust Sanger Institute, Wellcome Trust Genome Campus, Cambridge, UK}
\affil[7]{Centre for Tropical Medicine and Global Health, Nuffield Department of Clinical Medicine, University of Oxford, Oxford, UK}
\affil[8]{Mahidol-Oxford Tropical Medicine Research Unit, Bangkok, Thailand}
\affil[*]{\it cworby@hsph.harvard.edu}
\makeatletter
\renewcommand{\@biblabel}[1]{\quad#1.}
\makeatother
\date{}

\begin{document}
\maketitle

\begin{abstract}
Whole genome sequencing of pathogens from multiple hosts in an epidemic offers the potential to investigate who infected whom with unparalleled resolution, potentially yielding important insights into disease dynamics and the impact of control measures.  We considered disease outbreaks in a setting with dense genomic sampling, and formulated stochastic epidemic models to investigate person-to-person transmission, based on observed genomic and epidemiological data. We constructed models in which the genetic distance between sampled genotypes depends on the epidemiological relationship between the hosts. A data augmented Markov chain Monte Carlo algorithm was used to sample over the transmission trees, providing a posterior probability for any given transmission route. We investigated the predictive performance of our methodology using simulated data, demonstrating high sensitivity and specificity, particularly for rapidly mutating pathogens with low transmissibility. We then analyzed data collected during an outbreak of methicillin-resistant {\it Staphylococcus aureus} in a hospital, identifying probable transmission routes and estimating epidemiological parameters. Our approach overcomes limitations of previous methods, providing a framework with the flexibility to allow for unobserved infection times, multiple independent introductions of the pathogen, and within-host genetic diversity, as well as allowing forward simulation.
\end{abstract}

\section{Introduction}

A fundamental aim in the analysis of infectious disease epidemics is to identify who infected whom, however, achieving this is challenging, since transmission dynamics are generally unobserved. A probabilistic estimation of the transmission tree based on all available data offers many potential benefits. In particular, this can lead to improved understanding of transmission dynamics, provide a mechanism to quantify factors associated with heightened transmissibility and susceptibility to carriage and infection, and help identify effective interventions to reduce transmission. Pathogen typing can be used to cluster genetically similar isolate samples, which can rule out potential transmission routes. Whole genome sequence (WGS) data offers maximal discriminatory power through the identification of individual point mutations, or single nucleotide polymorphisms (SNPs), potentially leading to more accurate transmission tree reconstructions than hitherto possible. However, the joint analysis of genetic and surveillance data poses several challenges, as the relationship between epidemic and evolutionary dynamics is complex \cite{Ypma2013}. 

To date, genomic data have primarily been used to analyse transmission at a population rather than an individual level. This typically relies on a broad sample of individuals from a large population, with the aim of estimating past population dynamics over a long period of time. Phylogenetic analyses have been used to infer patterns of large-scale geographic spread \cite{Harris2010}. Coalescent theory has been used with such data to estimate, among other things, fluctuations in population size and transmission parameters \cite{Pybus2001,Volz2009}. Methods have also been described to estimate transmission parameters by combining sequence data and time series incidence data \cite{Rasmussen2011}. 

In contrast, we focus on individual-level transmission, using high-frequency genomic samples from a subpopulation (eg. hospital, school, jail, farm, community), with the aim of reconstructing transmission routes. Such sampling presents more of a challenge in terms of resources and data collection. However, with falling sequencing costs, gathering genomic data is rapidly becoming a feasible component of outbreak investigations, as demonstrated by recent studies \cite{Snitkin2012, Koser2012, Gardy2011}. We aim to estimate the transmission tree, a graph representing the spread of a pathogen between individuals, comprising nodes (cases, which may be defined as infected or colonized persons depending on the context), and directed edges (transmission events). Edges may additionally be associated with a transmission time. A transmission tree may be composed of multiple unconnected subtrees, each representing independent chains of transmission. Each transmission chain has an origin, representing a new introduction of the pathogen into the population. While in some situations, it may be reasonable to regard the tree as fully connected (that is, only one origin exists), more generally, multiple introductions of the pathogen from external sources must be accounted for.

A number of approaches to reconstruct transmission trees for communicable pathogens using densely sampled genomic data have been described in recent years. Many methods have been based around the construction of phylogenetic trees, which describe the inferred evolutionary relationships between pathogen samples, and can be fit to sequence data under various evolutionary models. The phylogenetic tree is a bifurcating structure in which external nodes represent sampled isolates, while internal nodes represent the most recent common ancestor of its descendants. Internal nodes are similarly linked, such that the structure is fully-connected. Since phylogenetic trees may be topologically dissimilar to transmission trees \cite{Pybus2009, Romero-Severson2014}, interpreting phylogenetic proximity as epidemiological linkage can be misleading. Furthermore, phylogenetic trees are undirected, leaving ambiguity around the direction of transmission even if the transmission tree is topologically identical.

Phylogenetic trees have been used in conjunction with contact tracing data using ad-hoc approaches to rule out possible transmission links \cite{Bryant2013, Gardy2011}, while other approaches have developed more formal methods to make use of phylogenetic trees to infer transmission trees. For instance, Ypma et al. developed a method to sample over both the transmission and phylogenetic tree given a set of sequence data, ensuring both structures remain consistent with one another \cite{Ypma2013}. This approach required the specification of a model to describe within-host pathogen dynamics, which remain poorly understood for the majority of pathogens. Similarly, Numminen et al. describe an importance sampling approach in which both phylogeny and transmission tree are sampled from proposal distributions \cite{Numminen2014}. This approach required sequence data to be partitioned into clusters pre-analysis, and the topology of the phylogeny to be fixed, but avoids the computational complexity associated with Markov chain Monte Carlo (MCMC) based methods.

Alternatively, a second class of reconstruction methods avoids phylogenetic tree inference, using models in which transmission routes are weighted by a function of observed genetic distance. Simply identifying the source of infection by selecting the host carrying the most genetically similar sampled isolate has been suggested \cite{Jombart2011}, although this neglects the role of within-host diversity and sampling time, as well as uncertainty surrounding the times of infection. While more sophisticated approaches allow for uncertainty in transmission time and provide a more realistic model for the accumulation of mutations over time, hosts are characterized by a single pathogen genotype \cite{Ypma2012, Morelli2012, Mollentze2014}. Jombart et al. describe a Bayesian data-augmentation approach making use of genetic distance data to infer likely transmission events, dates of infections, and unobserved cases \cite{Jombart2014}. The approach assumes known distributions of the generation interval and time from infection to isolate collection, and does not allow for within host diversity or explicitly account for imported cases (though multiple unconnected trees can be allowed for). These assumptions mean that, while the approach may be suitable for an acute infection in an outbreak scenario, it is not appropriate for pathogens such as {\it S. aureus} where long-term carriage is common, the generation interval is not well-defined, and where within-host diversity can be substantial.

Of the above methods, all assume that a single genotype is sampled from each host, with the exception of Numminen et al. \cite{Numminen2014}. This assumption can lead to poor tree inference in the presence of within-host diversity \cite{Worby2014a}. Only the approach developed by Mollentze et al. can identify importations \cite{Mollentze2014}; the remainder of methods assume the transmission tree is fully connected. Most methods described assume infection times are known with certainty. It is likely to be extremely useful to relax each of these assumptions in most infectious disease settings. Finally, while the importance sampling method by Numminen et al. can accommodate various transmission models \cite{Numminen2014}, the remainder consider instead the probability of a transmission tree linking the set of infected individuals, ignoring the probability of susceptible individuals avoiding infection. 

Here we describe a generalized approach to transmission tree reconstruction that overcomes these limitations and makes use of both molecular typing information and known exposure data. A key novelty of our approach is that we model the genetic distances between sequences rather than the microevolution of the sequences themselves. This offers a flexible framework in which multiple independent introductions of the pathogen and within-host diversity may be considered, as well as the transmission process itself. This approach avoids the need to make any assumptions about the within-host pathogen population dynamics, which in general, are poorly understood. Furthermore, our proposed framework allows data to be simulated forward in time, a feature lacking in the majority of existing methods (with reverse time simulation typically required in phylogenetic methods, and only an incomplete set of genetic distances simulated from other approaches), which is of fundamental importance in predictive modelling and model evaluation.

\section{Methods}

The importance of identifying transmission pathways in hospital epidemiology is one of the major motivations for our work. We therefore describe our approach for this setting, and analyse real and simulated hospital epidemic data. Since infection is often asymptomatic in this setting, even with frequent patient screening, epidemics are only partially observed. Furthermore, patients may be admitted to the ward already infected (importations), which requires consideration of multiple disconnected transmission trees. Our approach accounts for these complications. In line with most literature on hospital-associated infections, we subsequently use the term `colonized' to refer to patients who are either symptomatically or asymptomatically infected with the pathogen. 

We observe a set of $n$ patients admitted and discharged from a hospital over a study period. For each patient ($j$, say), we observe the day of admission $t^a_j$ and discharge $t^d_j$, the days and results of screening tests (positive or negative for the pathogen) taken during their stay. We denote the set of all screening results by $X$. We also suppose that some (not necessarily all) of the positive swabs have a corresponding sequenced isolate, i.e. we have genetic information related to some of the positive tests. From a total of $n_s$ sequenced isolates, we derive a symmetric pairwise genetic distance matrix $\Psi=(\psi_{a,b})_{a,b\leq n_s}$, with the genetic element $\psi_{a,b}$ giving the genetic distance between isolates $a$ and $b$. If colonized, the day of colonization for patient $j$ is denoted $t^c_j$, and the source of infection, $s_j$, is equal to the ID of the patient from whom the pathogen was acquired, or equal to zero if the patient was already colonized on admission. These quantities specify the transmission tree, but are typically unobserved. For patients who are never colonized, $t^c_j=s_j=\infty$. We denote the set of colonization times and routes of infection by $T$. We can write the likelihood of observing genetic and screening data, given model-specific parameters $\theta$ as

\begin{equation}
\pi(X,\Psi|\theta)=\int_T \pi(\Psi|X,T,\theta)\pi(X|T,\theta)\pi(T|\theta)\text{d}T
\end{equation}

We now describe the distinct components of our model, which govern the transmission dynamics ($\pi(T|\theta)$), the observation of screening data ($\pi(X|T,\theta)$), and the generation of genetic diversity $\pi(\Psi|X,T,\theta)$.

\subsection{Transmission model}

We first define a stochastic model which describes both pathogen transmission and the genetic distances arising between genotypes sampled from any two individuals. Each patient $j,\>j=1,\dots,n$, is admitted to the ward, independently carrying the pathogen with probability $p$, and has marker variable $\phi_j$, equal to 1 if the patient is positive on admission, and zero otherwise. We assumed homogeneous mixing, such that each colonized patient has equivalent contact with each susceptible individual. The rate of transmission to a given susceptible patient on day $t$ is then $\beta C(t)$, where $C(t)$ is the number of colonized patients on day $t$, and $\beta$ is the transmission rate per colonized individual. We assumed that individuals colonized on day $t$ may transmit the pathogen from day $t+1$ until their discharge. Working in discrete time using daily intervals, the probability that a given susceptible patient avoids colonization on day $t$ is $\exp(-\beta C(t))$, thus, acquisition occurs with probability $1-\exp(-\beta C(t))$. Each patient has the same chance of contacting any other patient in this model, and we note that transmission is often indirect, via the hands of healthcare workers (HCWs) \cite{Cooper1999,Pittet2008,Albrich2008}.
Given an individual acquires the pathogen on day $t$, the probability that the source of transmission is a particular positive individual is simply $1/C(t)$, since it is assumed that colonized patients have an equal potential to transmit. More generally, this probability will be the transmission pressure from the potential source divided by the total transmission pressure at time $t$. The model for transmission dynamics, $T$, can then be given as

\begin{eqnarray} 
\pi(T|\theta)&=&p^{\sum_j\phi_j} (1-p)^{n-\sum_j\phi_j}\\
&\times&\prod_{i=1}^n\left( \textbf{1}_{t^c_i=t^a_i}+\textbf{1}_{t^c_i\neq t^a_i}\exp\left\{-\sum_{t=t^a_i}^{\min(t^c_i-1,t^d_i)}\beta C(t)\right\}\right)\nonumber\\
&\times& \prod_{\substack{j:t^c_j\leq\infty,\\ \phi_j=0}}\left(\frac{1-\exp\{-\beta C(t^c_j)\}}{C(t^c_j)}\right),\nonumber
\end{eqnarray}

where $\textbf{1}_x$ is the indicator function, returning 1 if the condition $x$ is true, and zero otherwise.

\subsection{Observation model}

During each patient's stay in the hospital, regular screening is carried out to detect carriage of the pathogen. We assume that the test is highly specific, but imperfectly sensitive \--- that is, false positive results are not possible, but a positive patient is correctly screened positive with probability $z$ (test sensitivity) \cite{Perry2004}. Let $TP(X,T)$, $FN(X,T)$ and $FP(X,T)$ be the total number of true positive, false negative and false positive results in the screening data respectively, given the set of colonization times. The likelihood of observing the screening results, given test sensitivity and transmission times is

\begin{equation}
\pi(X|T,\theta)=z^{TP(X,T)}(1-z)^{FN(X,T)}\textbf{1}_{FP(X,T)=0}.
\end{equation}

\subsection{Genetic distance models}

We defined the genetic distance to be the observed number of SNPs between isolates, though other metrics are possible. The genetic distance between any two isolates is assumed to be drawn from some probability distribution, which in general can depend on any desired features of the two samples in question, or the hosts from whom they were sampled, such as their relatedness in terms of transmission. We assume that genetic distances are perfectly observed, and that insertions, deletions and recombinant sections are removed from the genome such that the genetic distance is representative of the accumulation of SNPs. 

The true distribution of the observed number of SNPs between two samples is complex, and depends on the mutation rate and the time of their most recent common ancestor, which in turn is dependent on the within-host pathogen population dynamics, as well as the effective transmission inoculum size. Since such factors are still poorly understood for most pathogens, we supposed that the distribution could be approximated by either a Poisson or a geometric distribution, dependent on the relationship between the sampled hosts. This relationship could be modeled a number of ways, but here we focus on two particular models, allowing for genetic diversity to be generated through alternative dynamics. 

\subsubsection{Transmission diversity model}
\label{sec:transdiv}
The first model, the transmission diversity model, discriminates between individuals in a transmission chain under the assumption that the expected genetic diversity changes predictably as sampled individuals are further apart in the tree. Typically, one would expect that distances will increase along the chain, due to the accumulation of mutations within each host. Each increase in the tree distance between nodes results in the expected genetic distance changing at a rate governed by a parameter $k$, which we call the transmission diversity factor. Distances between isolates taken from individuals in unrelated transmission chains are assumed to be drawn from a different specified distribution. 

We proposed a distribution to describe the genetic distance between two isolates, given the relationship between their carriers in the transmission tree. For isolates $x$ and $y$, we defined $t(x,y)$ to be the number of links which separate the isolates in the transmission tree, with $t(x,y)=\infty$ if $x$ and $y$ are sampled from separate chains. For two samples taken from the same host, we have $t(x,y)=0$. Under the transmission diversity model, we used the following geometric distribution: for $d=0,1,\dots$

\begin{equation}
\pi(\psi_{x,y}=d)=\left\{\begin{array}{cl}\gamma k^{t(x,y)}(1-\gamma k^{t(x,y)})^d&t(x,y)<\infty,\\
\gamma_G(1-\gamma_G)^d&t(x,y)=\infty,\\ \end{array}\right.
\end{equation}
where $\gamma k^{t(x,y)}\in[0,1]$. Here, the parameter $\gamma_G$ represents genetic diversity between samples belonging to different transmission chains. The parameter $\gamma$ is the geometric parameter for genetic distances occurring in the same transmission chain, while $k$ denotes the factor by which this parameter is changed upon an additional transmission link between the samples.

The expected genetic distance between samples is then $(1-\gamma k^z)/\gamma k^z$ for samples separated by $z$ transmission links, or $(1-\gamma_G)/\gamma_G$ for samples belonging to independent chains. The likelihood contribution for the $n$th observed sequence is then just the product of probabilities for the $n-1$ genetic distances to previously observed sequences. Under this model, the likelihood of observing the genetic distance matrix $\Psi$, given the transmission tree structure, is:

\begin{align}
\pi(\Psi|X,T,\theta)=\prod_{y=2}^{n_s}\prod_{x=1}^y&\left(\textbf{1}_{t(x,y)\leq\infty}\gamma k^{t(x,y)}(1-\gamma k^{t(x,y)})^{\psi_{x,y}}+\right.\\
&\left.\textbf{1}_{t(x,y)=\infty}\gamma_G(1-\gamma_G)^{\psi_{x,y}}\right)\nonumber
\end{align}

We note that in regular circumstances, we would expect $k\leq1$, indicating steady or increasing diversity along a transmission chain. However, we allow for $k$ to take values greater than 1, as this may highlight sampling bias (eg. hosts with greater within-host diversity sampled more frequently), which would not be revealed with a fixed upper bound of 1.

The true distribution of genetic distances between independent transmission chains is dependent on the population which enters the hospital already colonized. This distribution will depend upon the strain types circulating in the community, and may be multimodal, reflecting clusters of similar strains. In the absence of local and regional sampling data which would be necessary to obtain a more suitable approximation, we use the geometric distribution, assuming strains are more likely to be similar than dissimilar. Our second model is designed to avoid the challenge of approximating this distribution.

\subsubsection{Importation structure model}
The second model, the importation structure model, assumes that imported cases are assigned into genetically similar groups. An individual who acquires the pathogen from another person in a given group is assigned the same group. An importation may belong to a previously observed group, despite not being connected in the transmission chain. The distance between each pair of isolates in a particular group follows the same distribution, regardless of the tree distance between the nodes, while we expect that isolates belonging to different groups to be genetically further apart. The number, and composition, of groups is unobserved, so must be inferred. Under the importation structure model, we have, for $d=0,1,2,\dots$

\begin{equation}
\pi(\psi_{x,y}=d)=\left\{\begin{array}{cl}\gamma(1-\gamma)^d&x\text{ and }y\text{ in the same group},\\
\gamma_G(1-\gamma_G)^d&\text{otherwise}.\\ \end{array}\right.
\end{equation}

Similar to the previous model, the expected genetic distance between samples is then $(1-\gamma)/\gamma$ for samples within the same group, or $(1-\gamma_G)/\gamma_G$ for samples belonging to different groups. It is necessary to introduce some additional notation for this model; let $g_j$ be the group to which colonized individual $j$ belongs (equal to zero if not colonized). We estimate an additional parameter, $c$, which gives the probability that the strain of an imported case belongs to an existing group. Under this model, the likelihood of observing the genetic distance matrix $\Psi$, given the transmission tree structure and group memberships $g$, is

\begin{equation}
\pi(\Psi|X,T,g,\theta)=\prod_{y=2}^{n_s}\prod_{x=1}^y
\left(\textbf{1}_{g_i=g_j}\gamma(1-\gamma)^{\psi_{x,y}}+\textbf{1}_{g_x\neq g_y}\gamma_G(1-\gamma_G)^{\psi_{x,y}}\right).
\end{equation}

Furthermore, the likelihood of observing $n_c$ groupings among the $\sum_j\phi_j$ importations is
\begin{equation}
\pi(g|\theta)=c^{n_c}(1-c)^{\sum_j\phi_j-n_c}.
\end{equation}

\subsection{Inference methods}
\label{sec:inf_methods}
To allow for unobserved transmission dynamics, namely the set of transmission days and sources $T=\{t^c,s\}$ (and additionally the set of group memberships $g$ under the importation structure model), we used a Bayesian framework and employed a data augmented MCMC algorithm \cite{Tanner1987} to sample over this space. Individuals with no observed positive swabs may also have been colonized, and we allow for this possibility by sampling over this space. A combination of Metropolis-Hastings and Gibbs sampling was used to draw samples from the parameter space $\theta$, consisting of the parameters $\{p,z,\beta,\gamma,\gamma_G,k\}$ for the transmission diversity model, and additionally $c$ under the importation structure model. This approach is an extension of the analytical frameworks previously used to estimate transmission parameters given unobserved infection days \cite{Oneill1999, Worby2013, Kypraios2010}. In addition to sampling transmission days, we specify a model for genetic data in this approach,  sampling transmission routes to identify the posterior transmission tree.

Transmission trees were sampled by randomly drawing new colonization days and sources, such that every proposed tree had a non-zero likelihood. Full details of the tree sampling methods, acceptance probabilities and MCMC algorithm are provided in the Appendix. By calculating the proportion of total samples for which particular transmission routes existed, we derived a tree with edges weighted by posterior probability. The R package `bitrugs' (Bayesian Inference of Transmission Routes Using Genome Sequences) contains code to implement the MCMC algorithm, and is included in the online supplementary materials.

Except where mentioned, parameters $p$, $z$, $\gamma$, $\gamma_G$, $c$ were assigned Beta(1,1) prior densities. The parameters $\beta$ and $k$ were assumed to be exponentially distributed {\it a priori}, with rate $10^{-6}$. 

\begin{figure}[!ht]
\centering
\includegraphics[width=12cm]{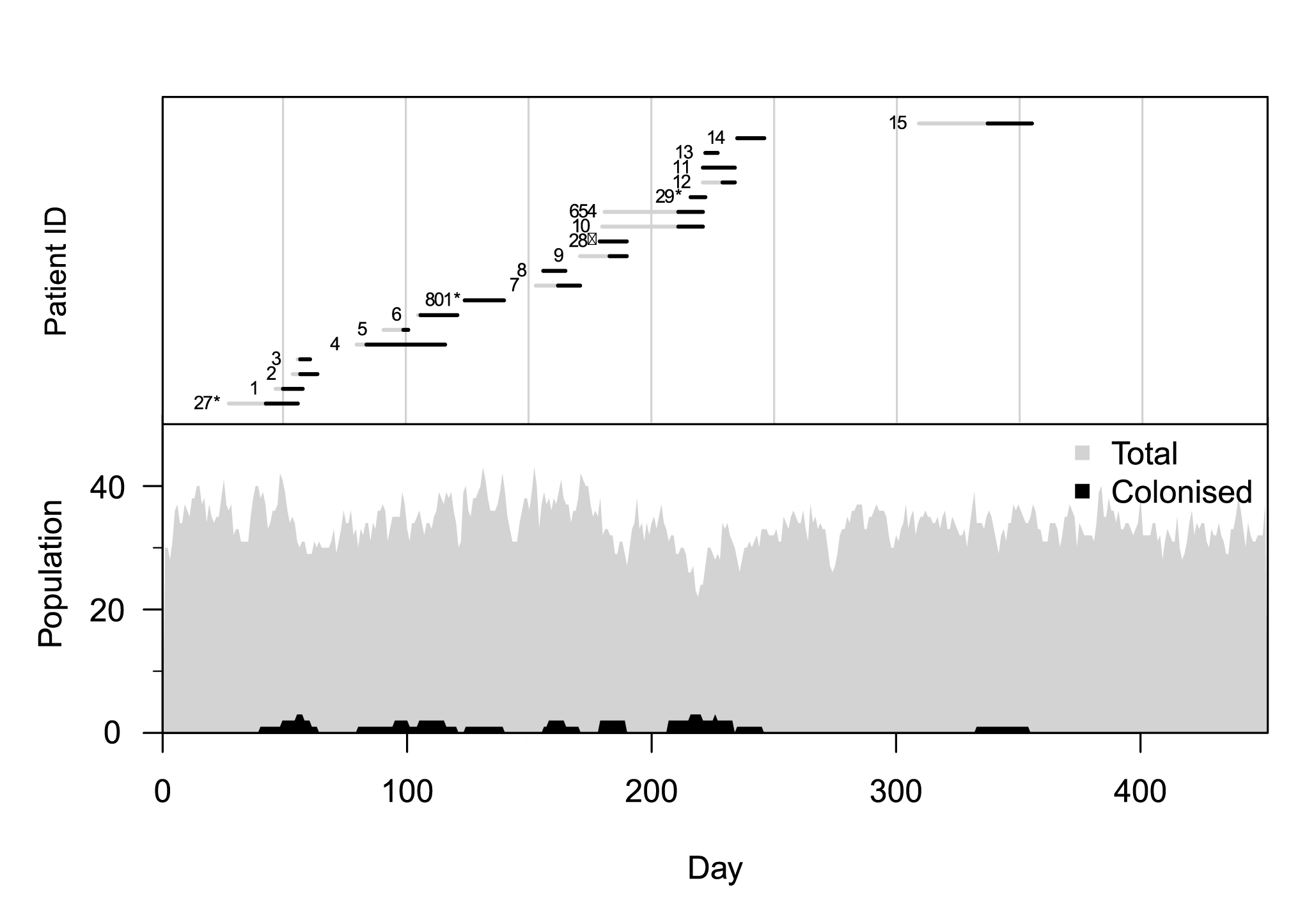}
\caption{Colonized patient episodes in the Rosie hospital neonatal ward. Patients are shown as colonized (black) after their first MRSA positive swab result until the end of their episode. Susceptible patients are shown in grey. Patient marked with an asterisk (*) carry a non-outbreak sequence type.}
\label{fig:data}
\end{figure}

\subsection{Data}

We first investigated the performance of our models using simulated hospital data, generated under several different scenarios. Code to simulate data is included in an R package, available in the supplementary materials. We assessed tree accuracy by comparing the simulated true and estimated tree, and examining the receiver operating characteristic (ROC) curve \cite{Krzanowski2009}, identifying scenarios in which the model performed well and poorly. We compared our estimated trees to the `uninformed' tree \--- that is, an estimate of transmission routes excluding genomic data, assigning each potential source an equal weight. The ROC for the uninformed tree is calculated under the assumption that the times of infection are known, an advantage over our estimation method. Calculating the area under the curve (AUC) and comparing this with the uninformed tree can indicate the improvement in accuracy over the na\"ive structure. 

We then applied our methods to methicillin-resistant {\it S. aureus} (MRSA) carriage and sequence data collected from a special care baby unit in Cambridge, UK, during an outbreak in 2011. These data comprised a full set of patient admission and discharge days, MRSA carriage screening results and sequenced genomes of a subset of positive results. The genomic data have been described previously by Harris et al., who combined genomic analysis and contact tracing to estimate routes of infection within and outside of the hospital ward \cite{Harris2013}.

\section{Results}

\begin{figure}[!ht]
\centering
\includegraphics[width=12.5cm]{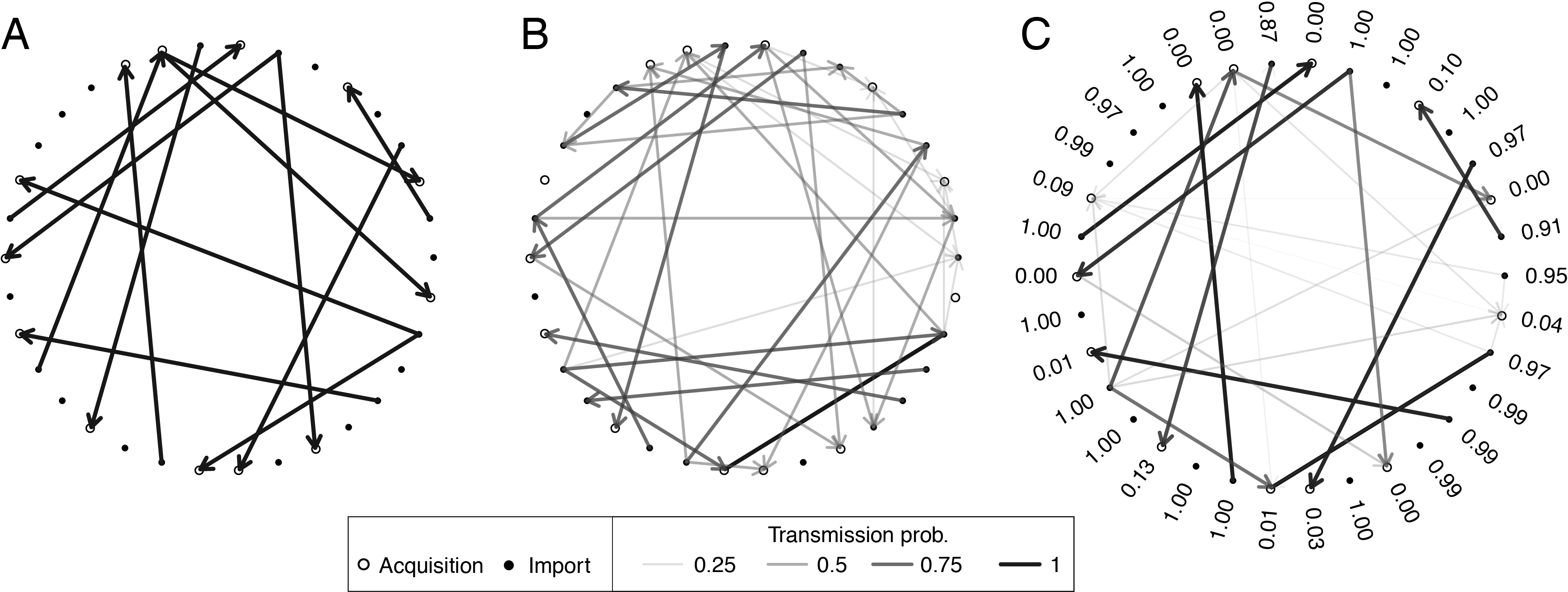}
\caption{A hospital outbreak was simulated, and we attempted to recover the routes of transmission. Patients are represented by open or closed circles, representing acquisitions or importations respectively, and transmission routes are shown as arrows. (A) The true transmission tree. (B) The uninformed transmission tree, in which all colonized patients at the time of transmission are considered equally likely sources of infection. (C) The estimated transmission tree under the transmission diversity model. Numbers beside each node represent the estimated probability that the individual was positive on admission.}
\label{fig_networks}
\end{figure}

\subsection{Simulated data}
\begin{figure}[!ht]
\centering
\includegraphics[width=8cm]{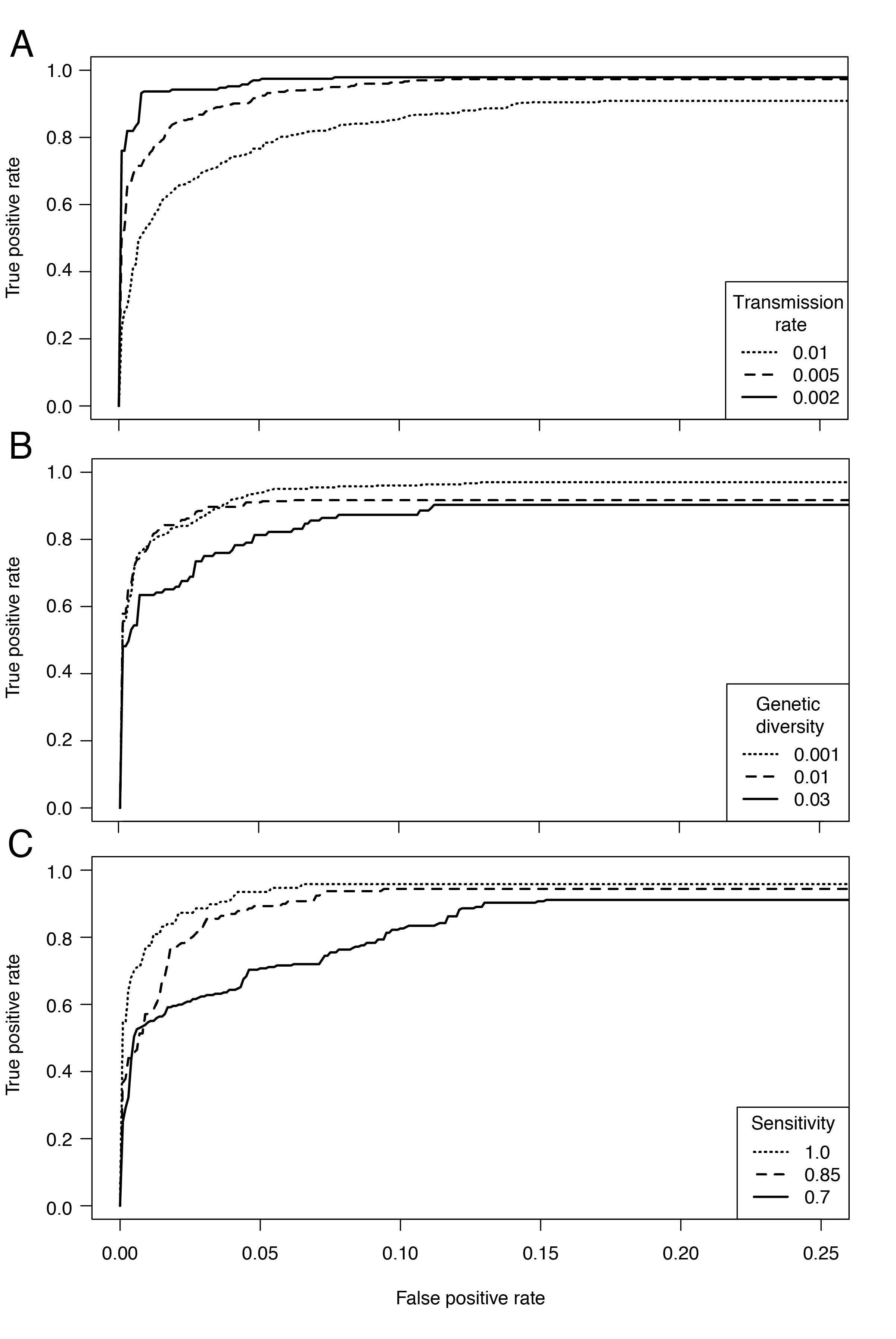}
\caption{ROC curves for estimated transmission trees, based on data simulated under various parameters. We varied transmission rate (A), the geometric rate parameter governing between-chain genetic diversity, for which lower values correspond to larger genetic distances (B), and test sensitivity (C). The ROC curves shown are the average for ten datasets simulated for each scenario.}
\label{fig_roc}
\end{figure}

We simulated several datasets under the two genetic distance models described in order to determine the ability of our estimation approach to recover the transmission tree as well as the parameter values. We simulated 500 patient admissions over 250 days, and varied model parameters to determine their impact on the ability to identify transmission routes (see supplementary material for more details on simulation). We also investigated the accuracy of tree reconstruction when fitting the model to data simulated under the alternative model. For a range of plausible parameter values we were able to recover the transmission tree well, consistently outperforming the uninformed transmission tree. Under both models, larger outbreaks tended to be associated with more uncertainty surrounding the source of infection. Figure \ref{fig_networks} shows a simulated hospital outbreak, comprising several unconnected subtrees. Also shown is the uninformed transmission tree, in which edges are placed with equal weight for all potential sources of transmission, and our reconstruction under the transmission diversity model. While most transmission events are successfully recovered, there is uncertainty within the largest transmission chains which contain several nodes, as well as, in some cases, uncertainty as to whether a case was imported or not. For simulations with an increased transmission rate, a higher number of genetically similar new infections were seen in the ward at any given time, increasing tree uncertainty (Figure \ref{fig_roc}A). The transmission diversity model allows the length of the transmission chain to have an impact on the expected genetic distances between two given isolates and therefore allows discrimination between the set of possible sources. For higher transmission rates, transmission chains typically become longer, resulting in the expected genetic distance between isolates approaching the levels expected for unrelated individuals, adding further between-chain uncertainty. Allowing the between-chain expected genetic distance to increase (i.e. reducing $\gamma_G$) resulted in improved accuracy (Figure \ref{fig_roc}B). If imported strains are always highly distinct, then it is straightforward to assign an individual to the correct chain, if not the true source of transmission. Table \ref{tab:scenarios} gives an overview of tree estimation accuracy under various parameter values.

\begin{table}[h]
\small
\begin{center}
\begin{tabular}{ll|cc}
Scenario&Parameters&AUC&AUC\\
&&(uninf.)&(inf.)\\
\hline
Baseline&$*$&0.67&{\bf 0.93}\\
Low sensitivity&$z=0.6$&0.67&{\bf 0.84}\\
High sensitivity&$z=0.9$&0.68&{\bf 0.94}\\
Low transmission&$\beta=0.001$&0.62&{\bf 0.96}\\
High transmission&$\beta=0.008$&0.74&{\bf 0.91}\\
Equal diversity ratio&$\gamma=0.1$,&0.68&{\bf 0.91}\\
&$\gamma_G=0.1$&\\
Low diversity ratio&$\gamma=0.3$,&0.68&{\bf 0.93}\\
&$\gamma_G=0.1$&\\
High diversity ratio&$\gamma=0.3$,&0.68&{\bf 0.96}\\
&$\gamma_G=0.005$&\\
No increasing&$k=1$&0.68&{\bf 0.93}\\
chain diversity&&\\
Strongly increasing&$k=0.5$&0.69&{\bf 0.90}\\
chain diversity&&\\
\hline
\end{tabular}
\caption{Estimated tree accuracy under various scenarios. Each value presented is the mean area under the ROC curve (AUC) for estimated trees under the transmission diversity model, based on 20 datasets simulated under the parameters indicated. Uninformed AUC is based on assigning equal weighting to all available sources. The more accurate reconstruction is highlighted in bold. $^*$Baseline scenario: $p=0.05$, $z=0.8$, $\beta=0.005$, $\gamma=0.2$, $\gamma_G=0.05$, $k=0.8$.
}
\label{tab:scenarios}
\end{center}
\end{table}

\begin{figure}[!ht]
\centering
\includegraphics[width=6cm]{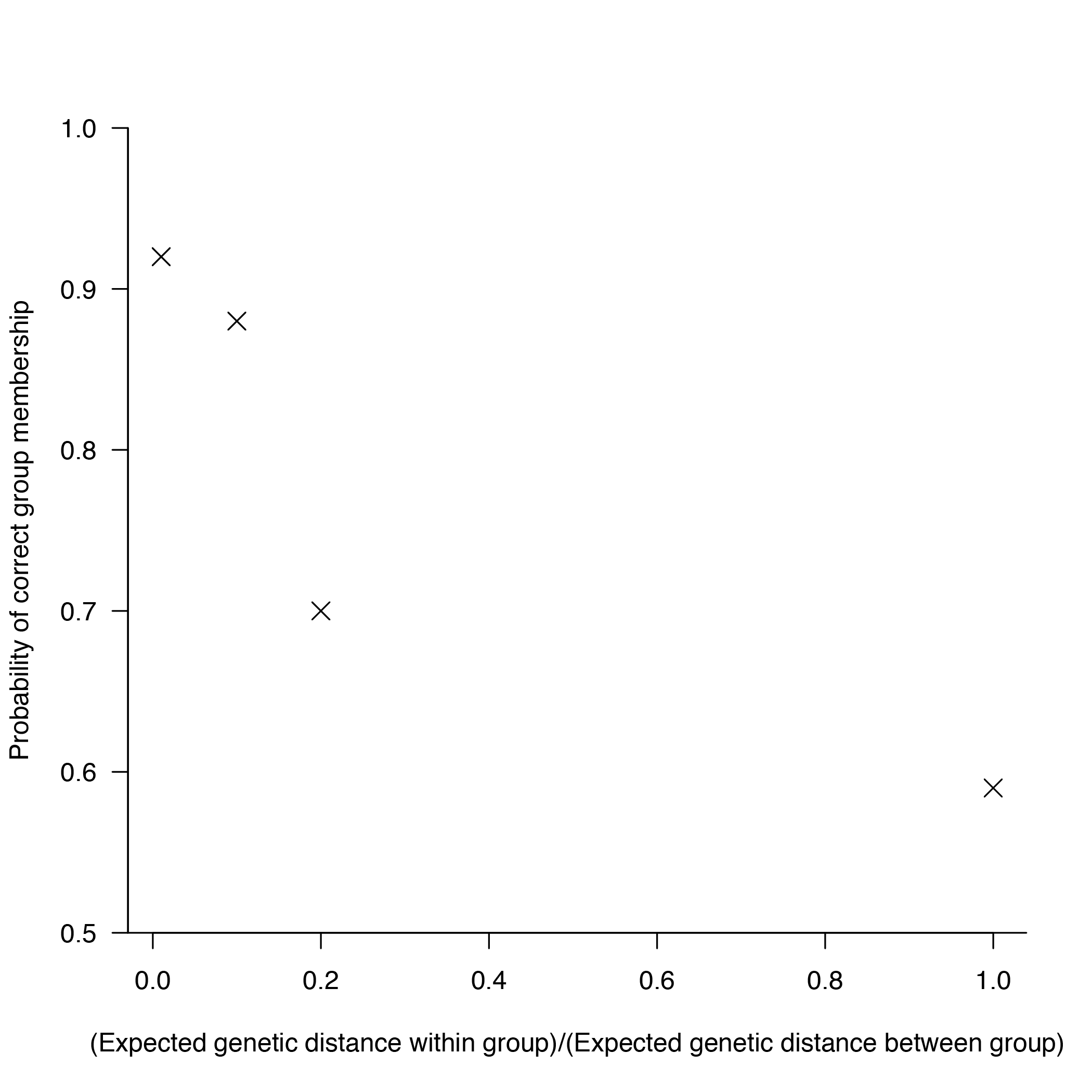}
\caption{Group identification under importation structure model. Data were simulated under a range of within and between group genetic distance distributions, and we estimated the posterior probability that the importation structure model placed an infected individual in the correct group (belonging to the same group as the first importation of that group). Baseline scenario: $p=0.05$, $z=0.8$, $\beta=0.005$, $c=0.2$.}
\label{fig:groups}
\end{figure}

The importation structure model lends itself to the identification of independent outbreaks rather than individual transmission routes, since by definition, it may discriminate between groups of similar strains, but assumes a fixed distribution of distances for all samples within a transmission chain. For this reason, tree reconstruction was often more uncertain than under the transmission diversity model, particularly for higher transmission rates. However, the identification of isolate groups was successful for a range of scenarios. In cases with frequent importations, the importation structure model often performed better than the transmission diversity model, particularly when importations were genetically similar to each other. Furthermore, this model generated better tree reconstruction from data simulated under the transmission diversity model, than vice-versa. The identification of group membership depended largely on the ratio of within- and between-group expected diversity; the smaller this value, the better the performance (Figure \ref{fig:groups}).


A key determinant of the transmission diversity model performance was the value of the factor $k$. The posterior estimate of this parameter was often associated with much uncertainty, especially in the absence of longer transmission chains. Differentiating the exact routes of transmission becomes difficult, or even impossible for values of $k$ close to 1, as genetic similarity along a transmission chain diminishes. Values of $k$ close to zero indicate that a considerable amount of mutation occurs between transmission events, and the genotype within the newly infected individual is very different to that found in the source. We found that tree reconstruction was less successful when $k$ was low (table \ref{tab:scenarios}), and low values of $k$ were typically overestimated. 

In most cases, the ROC curve for estimated transmission trees indicated a considerably better performance than the uninformed tree, demonstrating the gain in information associated with the inclusion of genomic data. However, the tree reconstruction was relatively poor where diversity was defined to be similar for related and unrelated isolates, or when diversity could accumulate quickly in a transmission chain (Table \ref{tab:scenarios}). Tree accuracy was relatively poor for lower values of test sensitivity (Figure \ref{fig_roc}C), but we nevertheless found that our estimates consistently outperformed the uninformed tree (Table \ref{tab:scenarios}). However, even with perfect sensitivity, some transmission routes were not recovered, due to colonization and subsequent discharge occurring prior to the next screening time. The degree of uncertainty surrounding even relatively simple trees is notable, reflecting the genetic similarity of linked cases.

We tested sensitivity to our choice of prior distributions by varying the rate parameter of the prior exponential distributions of $\beta$ and $k$. We found that neither the parameter estimates nor the estimated transmission tree were affected considerably by varying this value between $10^{-2}$ and $10^{-10}$.

We additionally simulated sequence data under an explicit pathogen evolutionary model. Using the R package `{\it seedy}' \cite{Worby2015}, we generated sequence data on top of transmission trees simulated as before. We found that transmission trees could be recovered well, offering a considerable improvement on the uninformed trees (see supplementary material for further details).

\subsection{MRSA outbreak data from Rosie Hospital, Cambridge, UK}

An outbreak of MRSA was observed in 2011 in a special care baby unit at the Rosie Hospital, Cambridge, UK, in which a total of 20 newborn infants were found to be MRSA-positive. We considered a dataset spanning 450 days, including this outbreak, comprising admission and discharge times, as well as MRSA screening results and times, for all patients admitted during this period. A total of 1108 unique patients were admitted to the ward in this period, and were swabbed regularly for the presence of MRSA. Figure \ref{fig:data} shows the colonized patient episodes and total population over the study period. Of the 20 patients with positive swabs, 18 had one positive isolate sequenced, and 15 of these were found to be sequence type 2371 (ST2371) (patient numbers 1-15). The remaining three sequenced isolates (carried by patients 27-29) were separated from this outbreak type (and each other) by several thousand SNPs. Two patients (654 and 801) had a positive swab, but no sequenced isolate. During the outbreak investigation, all HCWs were screened voluntarily and with consent, one of whom was found to be MRSA positive. Twenty colonies from this individual were sequenced, revealing carriage of several ST2371 genotypes, differing by up to 10 SNPs (mean pairwise distance 3.9 SNPs). Full details of sequencing and data collection are described in Harris et al. \cite{Harris2013}, and sequences were uploaded to the European Nucleotide Archive (\texttt{www.ebi.ac.uk/ena}). 

The non-outbreak sequence types differed by many thousands of SNPs. Fitting the transmission diversity model to these data using a geometric distribution would make the relative likelihood of an observed distance of a much smaller magnitude arising from unrelated transmission chains very low, forcing the model to link all outbreak strains where possible. This in turn results in an overestimation of the frequency of transmission events. This suggests that a geometric distribution is not an adequate approximation of between transmission chain genetic distances when multiple strain types are present. For this reason, we fitted the transmission diversity model to a restricted dataset, omitting the non-ST2371 strain types. Alternatively, a multimodal distribution could be chosen to account for distant strain clusters, although such a model would likely be overparameterized given the available data. The importation structure model avoids this issue, so we used all available data in this case. For both models, we assumed that test sensitivity was beta distributed with mean 0.8 and standard deviation 0.04 {\it a priori}, in line with previous estimates \cite{Worby2013}. We used the sequenced isolates from the colonized HCW to inform our prior density of within-host diversity, $\gamma$. All other priors were as described in section \ref{sec:inf_methods}.

\begin{table}[h]
\small
\begin{center}
\begin{tabular}{l|cc}
Parameter&Transmission diversity&Importation structure\\
&(95\% CrI)&(95\% CrI)\\
\hline
Probability of&0.012 (0.007, 0.019)&0.017 (0.009, 0.024)\\
importation, $p$&&\\
Test sensitivity, $z$&0.72 (0.65, 0.79)&0.70 (0.64, 0.77)\\
&&\\
Transmission rate&89.9 (38.8, 158.2)&80.6 (30.1, 153.7)\\
$\beta\times10^{-5}$&&\\
Within host/group&0.20 (0.18, 0.23)&0.22 (0.19,0.25)\\
diversity $\gamma$&&\\
Between host/group&0.17 (0.18, 0.23)&1.6 (1.4, 1.9)$\times10^{-4}$\\
diversity, $\gamma_G$&&\\
Chain diversity&1.2 (0.71, 1.82)&---\\
factor, $k$&&\\
\hline
\end{tabular}
\caption{Posterior mean estimates and 95\% credible intervals for parameters of each model fitted to the Rosie hospital outbreak data. 
}
\label{tab:mrsa}
\end{center}
\end{table}

\begin{figure}[!ht]
\centering
\includegraphics[width=14cm]{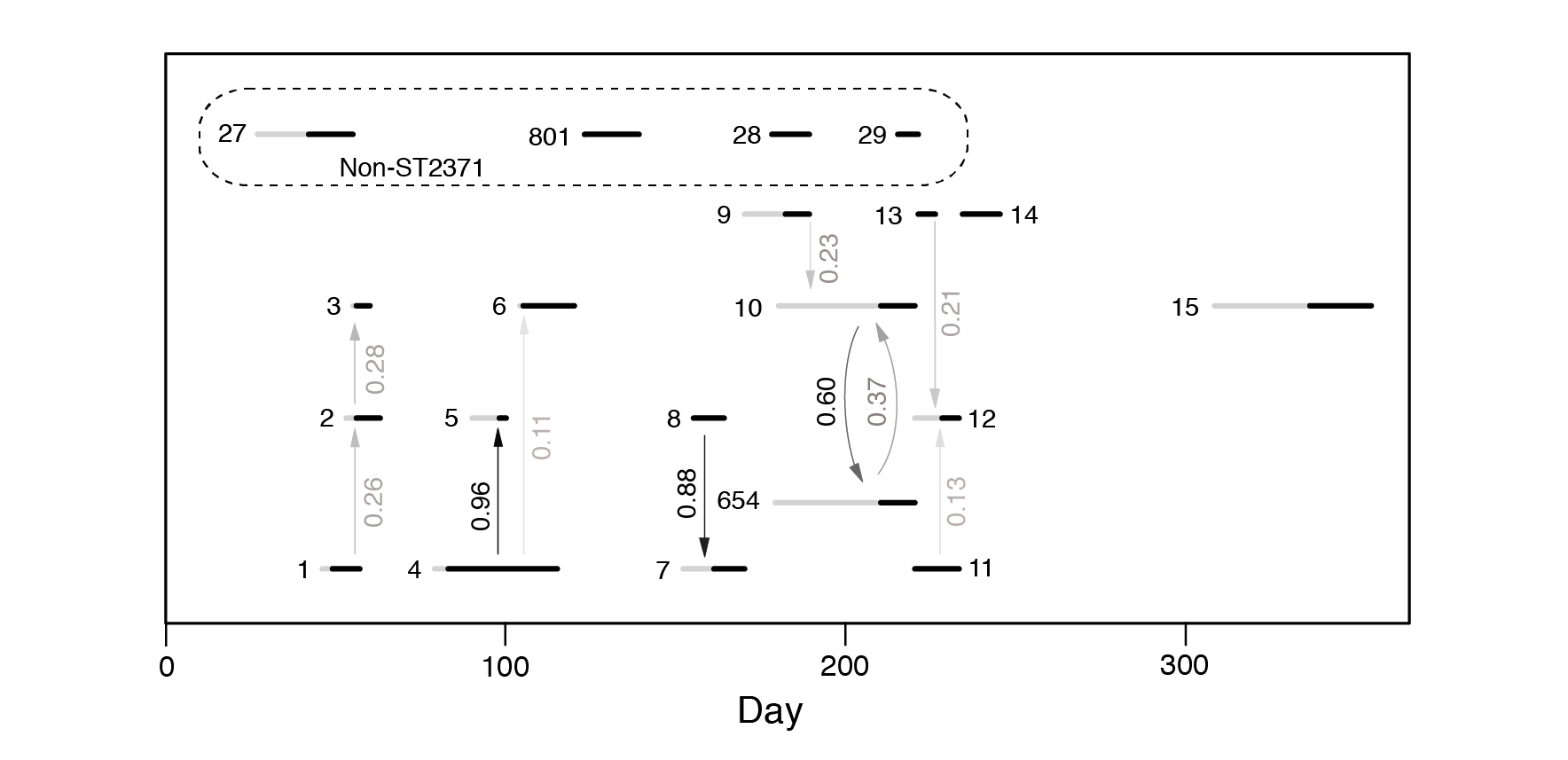}
\caption{Colonized patient episodes in the Rosie hospital neonatal ward. Horizontal bars represent patient episodes, with ID marked alongside. Grey bars denote susceptibility, while black represents the period after the patient's first MRSA positive swab. Arrows denote inferred routes of transmission, with darker arrows representing higher posterior probabilities, the values of which are given alongside.  Patients carrying non-outbreak types are shown at the top of the figure.}
\label{wardplot}
\end{figure}

\begin{figure}[!ht]
\centering
\includegraphics[width=12cm]{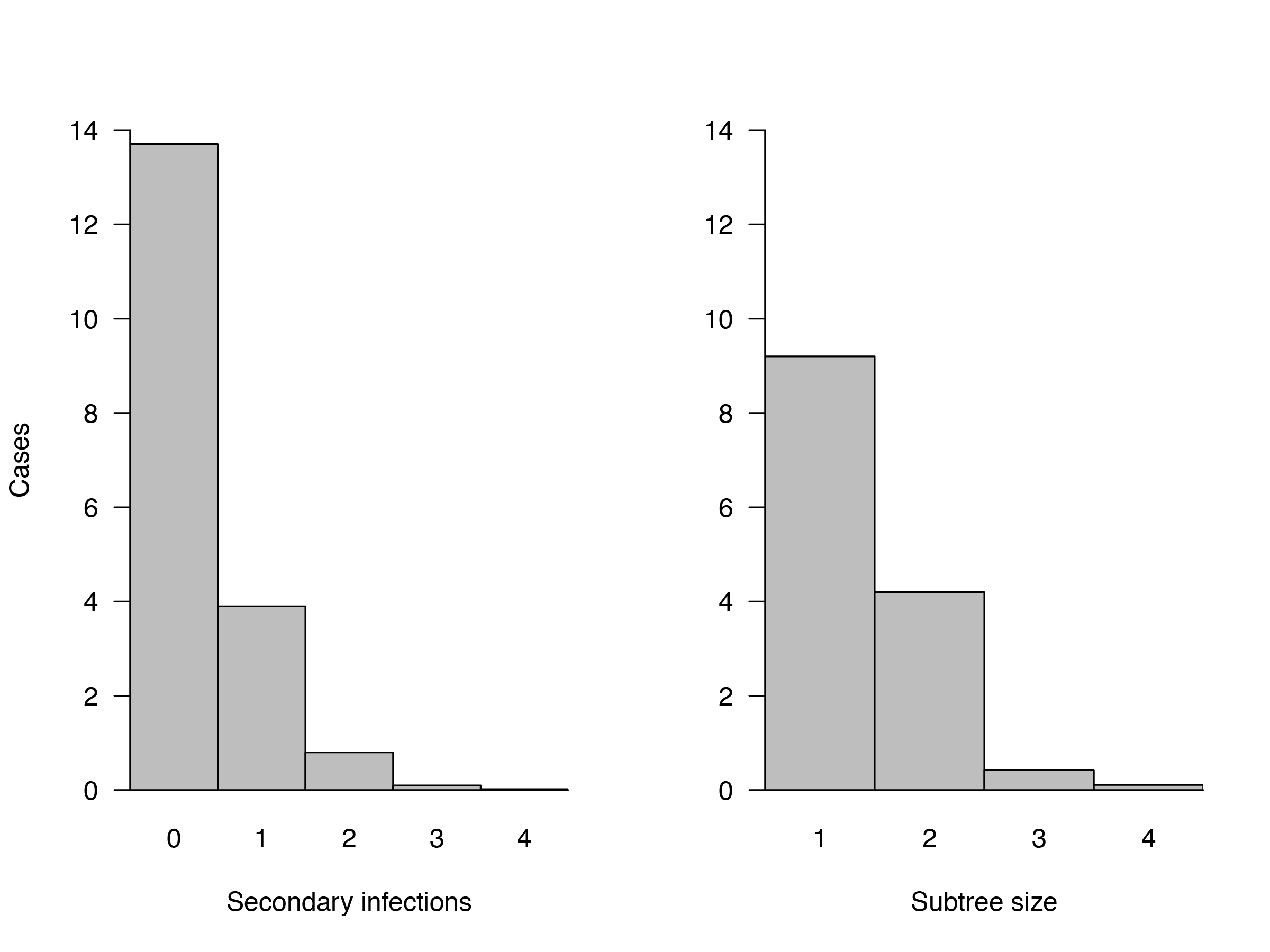}
\caption{Properties of the transmission network, estimated under the transmission diversity model. The posterior distribution of secondary infections for each colonized individual (left), and of the number of connected nodes in each subtree (right).}
\label{fig:chain}
\end{figure}

We first ran the MCMC algorithm under the transmission diversity model. Posterior mean estimates and credible intervals of model parameters are summarized in table \ref{tab:mrsa}. We estimated that 1.2\% (95\% CrI: 0.7\%, 1.9\%) of patients were positive on admission. The rate of transmission was low, and we estimated a total of 4.8 (3, 7) acquisitions on the ward. Three transmission events had a posterior probability above 0.5, and no transmission was inferred to or from the non-outbreak types (Figure \ref{wardplot}). Around 26\% of colonized individuals were the source of one or more secondary cases (Figure \ref{fig:chain}a). Isolates from patient 654 were not sequenced, therefore we sampled over possible genetic types for this individual. With a high posterior probability (97\%), this patient was involved in a transmission event with patient 10, although the direction of transmission was uncertain. We estimated the transmission diversity factor $k$ to be 1.2 (0.7, 1.8), the wide credible interval reflecting the paucity of transmission events, most of which formed a transmission chain of length 1 (Figure \ref{fig:chain}b). Within-host diversity was estimated to be 3.9 (3.3, 4.6) SNPs, an estimate dominated by the prior density based on the samples from the HCW. As such, the expected distance from source to recipient was approximately 3 SNPs. With the non-outbreak strain types excluded, the expected distance to unrelated strains was 4.9  (4, 6.1) SNPs. We generated the posterior predictive distributions for the number of observed importations, acquisitions and overall diversity. We found that the true observed values from the dataset fell within the 95\% central quantile of the predictive distribution, providing no indication that the model was a poor fit (see Appendix).

The importation structure model placed a high posterior probability on the existence of four groups, reflecting the four sequence types observed in the study. We estimated the expected pairwise distance between isolates belonging to the same group to be 3.7 (3, 4.5) SNPs. Under this model, the probability of importation was estimated to be slightly higher, while the transmission rate was lower. We estimated that patients 1 and 3, who were originally missed by the infection control team at the hospital, were part of the main outbreak group, in accordance with the previous study \cite{Harris2013}.

\section{Discussion}

The genetic diversity and structured importation models we have described here allow the combined analysis of genetic and epidemiological data. We applied these methods to the transmission of MRSA in hospitals, demonstrating the simultaneous estimation of model parameters and a transmission tree. More generally, the approaches we have developed can be applied to the analysis of disease transmission in a community where high-frequency sampling of sequence data is available. These methods offer flexibility not available in previous approaches, as they allow multiple introductions of the pathogen into the population, incorporation of within-host genetic diversity, unobserved colonization times, and the provision of estimates of uncertainty for each potential transmission route. While we have used whole genome sequence data, this approach may also be used with lower resolution genetic data, provided a distance metric between isolates can be defined. A major advantage of our framework over existing methods is the ability to simulate forward from our models. This allows one to perform predictive analyses, as well as model evaluation procedures. 

A considerable degree of uncertainty was associated with the resulting estimated transmission trees, even for small outbreaks, despite the densely sampled genomic data and well-defined periods of potential contact. As has been previously demonstrated, individual transmission routes are generally unlikely to be identified with high confidence using genetic distance data alone \cite{Worby2014}. This reflects the high genetic similarity of individuals in the same transmission chain, and we believe that quantification of uncertainty is of much importance – methods which provide an optimal tree with no measure of uncertainty may be misleading. While we have demonstrated the general improvement in tree accuracy associated with the availability of genomic data, in most cases, much uncertainty is likely to remain regarding transmission routes.

Some previous studies aiming to reconstruct transmission trees using densely-sampled genetic data have used a phylogenetic approach, implicitly assuming that a transmission tree will map closely to the phylogenetic tree \cite{Gardy2011,Bryant2013,Cottam2008}. However, this assumption may not hold \cite{Pybus2009,Romero-Severson2014}. A fundamental limitation of phylogeny-based approaches is that the relationship between the transmission and phylogenetic trees depends on the within-host evolutionary dynamics which, in the absence of dense within-host sampling, are not identifiable. By simultaneously sampling over the phylogenetic tree and the transmission tree, one can account for unknown coalescent times and dependencies between genetic distances \cite{Ypma2013}. While this approach offers a more realistic model for the emergence of diversity, it also requires a reliable model of within-host pathogen population dynamics. Furthermore, this method requires sampling over the space of phylogenetic trees (and therefore unobserved sequence data), resulting in a considerably more computationally intensive approach than our proposed framework. Even with such a model, the method cannot differentiate between importations and acquisitions, crucial when considering an outbreak in a hospital setting. Data on within-host dynamics are currently scarce, and these dynamics may vary widely between individuals. As such, robust specification of such models is challenging. 

Our analysis has some limitations. We have assumed that the source of transmission for each patient must come (indirectly, via HCW) from another patient present on the ward. As previously suggested, there is a strong possibility of external sources of transmission in this setting \cite{Harris2013}. This would mean that the patient-to-patient transmission rate may be overestimated in our model. Our approach would perform best when all potential contacts are included in the analysis. Additionally, we have used a transmission model that does not allow for heterogeneous rates of transmissibility. We believe that this model is adequate in this setting, and did not affect our primary goal of estimating the transmission tree. We have assumed that clearance of carriage and reinfection are not possible; while it appears unlikely that such events are common in this dataset, incorporating mechanisms for these could be important in other settings and over longer time periods. 

Our estimates from the Rosie hospital data suggested that within- and between-host diversity were similar, with the former slightly higher than the latter, suggested by the estimate of $k>1$.  Our estimates of within-host diversity were driven by the HCW, since multiple isolates were not collected from patients. If the HCW was colonized for a long period of time, a higher level of within-host diversity would be expected than within newly colonized infant patients, potentially leading to estimates of $k>1$. We believe that repeated sampling of each patient would lead to an improved estimate of within-host patient diversity, and that as an estimate of $k>1$ would be unlikely. We repeated our analysis with $k$ restricted to the interval $[0,1]$, and found that both parameter estimates and the inferred transmission tree were largely unaffected (supplementary material, table 2).

We chose simple geometric distributions to represent the genetic diversity both within and between individuals, assuming the probability of each observed sequence was time-homogeneous. We additionally experimented with equivalent Poisson distributions, however, results for the ICU data were very similar using both distributions, although this may not hold for larger datasets with longer transmission chains. While little evidence exists on observed genetic diversity during an epidemic, pairwise genetic distances of the same strain type collected during a tuberculosis outbreak appear to approximate a geometric distribution \cite{Bryant2013}, and with a known time to coalescence $t$ and mutation rate $\mu$, the genetic distance should follow a $\text{Pois}(2\mu t)$ distribution. With an unknown coalescent time and constant pathogen population size, the genetic distance between contemporaneously sampled genomes should follow a Geometric distribution \cite{Watterson1975}. 

As discussed in section \ref{sec:transdiv}, the true distribution between independent transmission chains may be multimodal, and poorly approximated by a geometric distribution. For this reason, we excluded non-outbreak sequence types manually before running the transmission diversity model. Our model could be extended in the future to remove this requirement, as for large datasets, and scenarios with concurrent outbreaks belonging to different strain types, this approach would be inappropriate. If local sampling data were available from the community or other regional hospital admissions, we could potentially construct an empirical distribution for the pairwise genetic distances expected between unlinked cases. As yet, such data are typically unavailable, though collection of such data may be feasible in the future.

We have assumed in our analysis that genetic distances are observed without error. In common with all existing tree estimation methods, we assumed that errors arising from sequencing and/or alignment were negligible. In the supplementary material we explored the impact of introducing observation error into the genetic distance matrix, finding that network reconstruction remained largely unaffected by such errors (see Appendix).

There are several potential alterations to our model which could be considered, and readily incorporated into our framework. The transmission chain diversity model allows the expected genetic distance to increase with number of transmission events, and could be reformulated to allow distance to increase linearly, or via an alternative relationship. Time between samples could instead be used as the factor by which diversity increases, however, this relationship is complex, and only fully understood by accounting for within-host dynamics \cite{Worby2014}. Furthermore, since the time between samples from transmission pairs does not vary greatly in this setting, we do not believe it would affect the results significantly. However, in cases where the length of stay (or length of carriage in a non-hospital setting) is long, which would allow times between sample pairs to vary considerably, then such an amendment should be considered.

Furthermore, in creating this model framework, we have assumed that genetic distances are drawn independently, which is not the case in reality. Although in principle this assumption can be relaxed, this would require considerable additional computational complexity. This may be considered in future studies.

Identifying imported cases is challenging, especially when cases are admitted with highly similar strains. In such a setting, our models can exhibit significantly different results \--- under the importation structure model, an importation of the same group is more likely than an acquisition soon after admission from another individual on the ward, while under the transmission diversity model, the reverse is true. As such, when strains circulating in the community are very similar to those found in the hospital, the importation structure model will generally perform better, allowing such strains to be clustered importations rather than rapid acquisitions. An intensive care unit admitting patients from elsewhere in the same hospital is an example of a setting where similar strains may be repeatedly imported to the ward. With no prior knowledge of external diversity, it is hard to determine which model is more suitable for identifying importations. However, if both models are run, significant differences between estimated transmission trees suggests that external diversity is similar to that found within the ward. Further data collection would be required to confirm this. The classification of cases as importations or acquisitions is key to the evaluation of infection control procedures, which for healthcare facilities in particular is of great importance. The framework described here can be used to provide evidence towards importation or acquisition in each case using genetic and surveillance data.

\section*{Acknowledgements}
We are grateful to Simon Harris, Est\'ee T\"or\"ok, Amanda Ogilvy-Stuart, Nick Brown and Cheryl Trundle for assistance with the Rosie Hospital data, as well as the anonymous referee and associate editor who provided insightful and constructive comments during the review process. This work was supported by funding from the European Community (Mastering Hospital Antimicrobial Resistance (MOSAR) network contract LSHP-CT-2007-037941). CJW received support from the National Institute of General Medical Sciences of the National Institutes of Health under award number U54GM088558. SJP received support from UKCRC Translational Infection Research Initiative (MRC grant number G1000803), and Public Health England. BSC was supported by The Medical Research Council and Department for International Development (grant number MR/K006924/1). The Mahidol Oxford Tropical Medicine Research Unit is part of the Wellcome Trust Major Overseas Programme in SE Asia (grant number 106698/Z/14/Z). The funders had no role in study design, data collection and analysis, decision to publish, or preparation of the manuscript.

\bibliography{refs}

\newpage

\section*{Appendix}
\subsection*{Data augmentation}

Since the full transmission process is typically unobserved, a data-augmented MCMC process was used. We sampled over the time of colonisation, as well as the source of colonisation, for each patient. Patients with no positive test results may have a colonisation time and source added (and subsequently removed) during the sampling process. Depending on assumptions made in the model, it may additionally be necessary to sample over the genetic distances arising from positive but unsampled hosts. In order to do this, one `phantom observation' is created for this individual, creating a new row (or column) of the genetic distance matrix $\Psi$, which we denote $\Psi^c_j$, when we propose to add a colonisation. This incorporates the uncertainty of unobserved colonisations to estimates of genetic diversity ($\gamma$ and $\gamma_G$). Probability mass functions $m(\cdot)$ and $m_G(\cdot)$ are defined, which are used to generate distances from this imputed sequence to isolates in the same group, and different groups, respectively. Since in our analyses, we have assumed that genetic distances are sampled independently, it is not necessary to create genetic observations for unsampled hosts. However, we describe the full process in this section. 

We describe here the data augmentation step for the importation structure model, where the genetic distance between strains depends on their assigned type. Due to the need to classify importations by MRSA type ($g$), the data augmentation step is more complex than for the transmission chain diversity model. The aim of the data augmentation process is to sample over the set of missing data $T=\{s,g,t^c,\phi,\Psi^c\}$, that is, the set of sources $s$, MRSA groups $g$, colonisation times $t^c$, admission statuses $\phi$, and a set of unobserved genetic distances, $\Psi^c$. Further, we define $Y_\text{ext}(t)=\{y_{i,1}:t^a_i<t,\, s_i=0\}$ to be the set of observed imported sequences prior to time $t$. 

At each iteration, a new dataset $T^*=\{s^*, g^*, t^{c*},\phi^*,\Psi^{c*}\}$ is proposed. Any patient who has a colonisation added by the algorithm is assigned a colonisation time and source, and a set of genetic distances from all other observed and inferred isolates. Let $v_s$ be the number of patients never screened positive, $v_q$ be the number of patients who carry MRSA at some point during their episode (either observed, or added by the algorithm), $v_a$ be the number of patients for whom a colonisation time has been added by the algorithm, $v_0$ of whom have no `offspring'; that is, the inferred colonised patients who infect no further individuals. Finally, let $v_n$ be the number of patients who have a positive screen, but no sequenced isolates. We define the proposal ratio $q_{A,A^*}=P(T^*\rightarrow T)/P(T\rightarrow T^*)$. At each iteration of the algorithm, one of the following moves is made with equal probability:

\begin{itemize}
\item {\bf Change colonisation route/time.} Select uniformly at random one of the $v_q$ patients ($j$, say) with a colonisation time. If $v_q=0$, no move is made. With probability $w$, propose the patient was positive on admission ($\phi_j^*=1$), otherwise sample a colonisation time $t^{c*}_j$ from $\{t_j^a,\dots,l_j\}$, where $l_j$ is the last potential day of colonisation (the earliest from day of discharge, day of first positive screen, and first onward transmission). If an importation is proposed, then with probability $w'$, we set $g^*_j$ to the same group of one of the $Y_\text{ext}(t^a_j)$ already-observed imported patients, otherwise, set $g_j^*=j$. If an acquisition has been proposed, we then select one of the $C(t^{c*}_j)$ patients already colonised on the proposed transmission day (excluding the chosen patient, if present on day $t^{c*}_j$) to be the source of colonisation. If there are no other colonised patients on this day, the move is rejected. We define $q_{T,T^*}$ according to the following table, where the row denotes the current state, and the column is the proposed state:
$$
\begin{array}{l|ccc}
&\text{Acquisition}&\text{Importation}&\text{Importation}\\
&&(g^*_j\neq j)&(g^*_j=j)\\
\hline
\text{Acquisition}&\frac{C(t^{c*}_j)}{C(t^c_j)}&\frac{|Y_\text{ext}(t^a_j)|(1-w)}{ww'(l_j-t_j^a+1)C(t^c_j)}&\frac{1-w}{w(1-w')(l_j-t_j^a+1)C(t^c_j)}\\
\text{Importation}&\frac{ww'(l_j-t^a_j+1)C(t^{c*}_j)}{|Y_\text{ext}(t^a_j)|(1-w)}&1&\frac{w'}{|Y_\text{ext}(t^a_j)|(1-w')}\\
(g_j\neq j)&&&\\
\text{Importation}&\frac{w(1-w')(l_j-t_j^a+1)C(t^{c*}_j)}{1-w}& \frac{(1-w')|Y_\text{ext}(t^a_j)|}{w'}&1\\
(g_j=j)&&&
\end{array}
$$

\item {\bf Change genetic distances.} Select one of the $v_n$ individuals with a positive screen, but no genetic data ($j$, say). If $v_n=0$, no move is made. Update their set of $n_s+v_a$ genetic distances $\Psi^{c*}_{j,1},\dots,\Psi^{c*}_{j,n_s+v_a}$. These distances are drawn at random according to the probability mass function $m$ and $m_G$ if the sequence being compared is taken from a related or unrelated chain respectively. This move has proposal ratio $$q_{T,T^*}=\frac{\prod_{i\neq j} (\textbf{1}_{g_i=g_j}m(\Psi^c_{j,i})+ \textbf{1}_{g_i\neq g_j}m_G(\Psi^c_{j,i}))}{\prod_{i\neq j} (\textbf{1}_{g_i=g_j}m(\Psi^{c*}_{j,i})+ \textbf{1}_{g_i\neq g_j}m_G(\Psi^{c*}_{j,i}))}.$$

\item {\bf Add colonisation.} Select at random one of the $v_s-v_a$ patients ($j$, say) who is currently assumed to be negative. If $v_s-v_a=0$, no move is made. With probability $w$, define this patient to be an importation, otherwise, an acquisition. If an importation is proposed, set $\phi^*_j=1$, $t^{c*}_j=t^a_j$. Now, we determine whether the proposed importation is clustered (in which case a group must be chosen) or not. With probability $w'$, propose the sequence is clustered, and select at random one of the already-observed imported sequences $Y_\text{ext}(t^a_j)$, setting the proposed MRSA group $g^*_j$ to that of the chosen sequence. If $|Y_\text{ext}(t^a_j)|=0$, the move is rejected. Draw a set of $n_s+v_a$ genetic distances $\Psi^{c*}_{j,1},\dots,\Psi^{c*}_{j,n_s+v_a}$ from probability mass functions $m(\cdot)$ and $m_G(\cdot)$, for strains in the same group and different groups respectively.

With probability $1-w'$, the sequence is not clustered, so the chosen individual is assigned to a new group; $g^*_j=j$. Draw a set of $n_s+v_a$ genetic distances $\Psi^{c*}_{j,1},\dots,\Psi^{c*}_{j,n_s+v_a}$ from the probability mass functions $m_G(\cdot)$ to all other sequences.

If an acquisition is proposed, then draw a colonisation time $t^{c*}_j$ from $\{t^a_j,\dots,t^d_j\}$. Select with equal probability a transmission source $s_j^*$ from the $C(t^{c*}_j)$ colonised patients on that day. If there are no colonised patients on this day, no move is made. Finally, select a set of $n_s+v_a$ genetic distances, according to the relationship between the chosen patient and other colonised patients. 

\item {\bf Remove colonisation.} Choose at random one of the $v_0$ patients who have had a colonisation time added by the data augmentation process, and are not currently assumed to be the source of infection for another individual. If $v_0=0$, then no move is made. Set $\phi^*_j=0$, $t^{c*}_j=\infty$, $g^*_j=0$ and $s_j^*=0$.
\end{itemize}

Having established the augmented data move mechanisms, the probability ratios $q_{T,T^*}$ for adding or removing colonisation times may be given as follows:

$$
\begin{array}{l|ccc}
&\text{Importation}&\text{Importation}&\text{Acquisition}\\
&\text{(clustered)}&\text{(unclustered)}&\\
\hline
\text{Add}&\frac{(v_s-v_a)|Y_\text{ext}(t^a_j)|}{ww'(v_0+1)M_a}&\frac{v_s-v_a}{w(1-w')(v_0+1)M_a}&\frac{(v_s-v_a)(t^d_j-t^a_j+1)C(t^{c*}_j)}{(1-w)(v_0+1)M_a}\\
\text{Remove}&\frac{ww'v_0M_r}{(v_s-v_a+1)|Y_\text{ext}(t^c_j)|}&\frac{w(1-w')v_0M_r}{v_s-v_a+1}&\frac{v_0(1-w)M_r}{(t_j^d-t_j^a+1)(v_s-v_a+1)(C(t^c_j)-1)}\\
\end{array}
$$
where 
$$M_a=\prod_{i=1}^{n_s+v_a} (\textbf{1}_{g_i=g^*_j}m(\Psi^{c*}_{j,i})+ \textbf{1}_{g_i\neq g^*_j}m_G(\Psi^{c*}_{j,i})),$$
and
$$M_r=\prod_{j:i\neq j} (\textbf{1}_{g_i=g_j}m_C(\Psi^{c*}_{j,i})+ \textbf{1}_{g_i\neq g_j}m_G(\Psi^{c*}_{j,i})).$$

Having sampled a candidate colonisation time/source, the candidate augmented dataset $T^*$ is accepted with probability
$$\text{min}\left(1,\frac{\pi(X,\Psi|T^*,\theta)\pi(T^*|\theta)}{\pi(X,\Psi|T,\theta)\pi(T|\theta)}q_{T,T^*}\right).$$

The proposal probability mass functions $m$ and $m_G$, which are used to generate unobserved sequences related to a transmission source, an external imported strain, or the reference strain respectively, should be specified pre-analysis. Similarly, one must set $w$ and $w'$, the probabilities of selecting an importation, and choosing an importation cluster. These choices will not affect results, but will impact the convergence and mixing rates of the algorithm.

Performing this process over a large number of iterations will allow us to calculate the posterior probability that a particular transmission route exists; this can be calculated as the proportion of iterations for which an inferred route is made.	

The data augmentation process is implemented similarly for the transmission chain diversity model. The same moves are proposed, but the imputation of groupings, $g$, is not required. For reasons of brevity, we omit the full description of the data augmentation process for the transmission chain diversity model.

\begin{figure}[!ht]
\centering
\includegraphics[width=8cm]{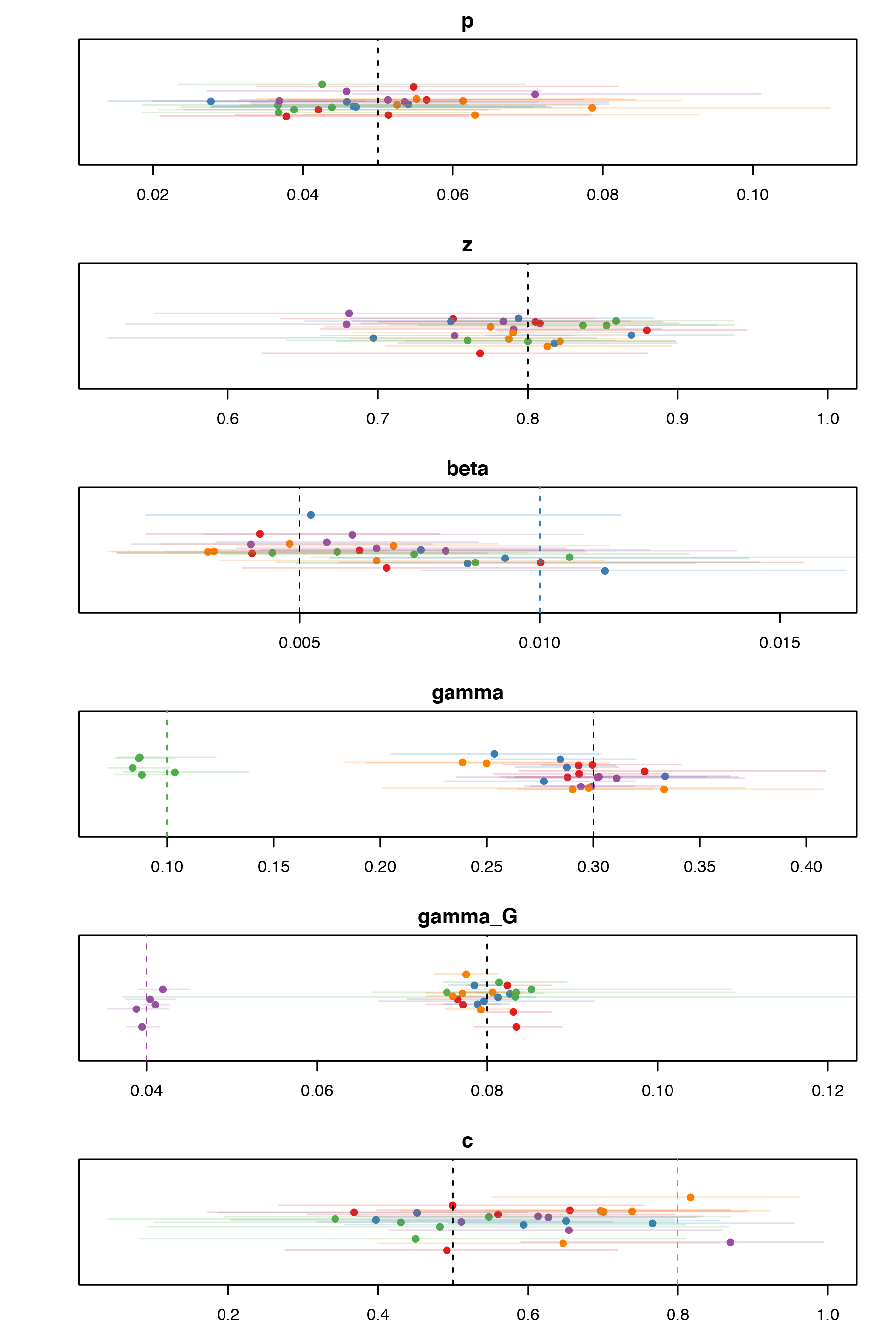}
\caption{Estimated parameters for the importation cluster model for datasets simulated under various scenarios. Each panel shows posterior median estimates and 95\% credible intervals for each parameter. Colours represent the scenario under which the data were simulated; baseline (red), high transmission (blue), high within-host diversity (green), high between-chain diversity (purple), and high group clustering parameter (orange). True values are shown as dashed vertical lines.
}
\label{fig:sim_recover1}
\end{figure}

\subsection*{Simulations}
\label{sec:sims}
In order to assess the performance of our model, we simulated epidemiological and genetic data for hospital wards according to each model. We now describe in detail how data may be simulated under either of the models described. Patient episodes are generated with probability $p$ of carriage on admission, and a length of stay is drawn from a Poisson distribution with mean $D$. Tests are generated every $x$ calendar days, and positive patients are observed to be negative with probability $1-z$. Patients positive on admission are assigned a set of genetic distances to all previously observed sequences (if applicable), which are drawn from distributions according to the relationship between isolates. For the transmission chain diversity model, genetic distances are generated by randomly drawing samples from a Geom$(\gamma_G)$ distribution. For the importation structure model, an importation sequence is defined to be unclustered if no previous importation sequences have been recorded. If the sequence is not the first to be observed, the strain is defined to be clustered with probability $c$, otherwise, it is unclustered. For genetic distances to isolates of the same type, we draw genetic distances at random according to the distribution Geom$(\gamma)$, while for sequences in a different group, genetic distances are drawn from the Geom$(\gamma_G)$ distribution.

Susceptible patients become colonized at a rate of $\beta C(t)$ at time $t$. Colonized patients contribute to the colonized population $C(t)$ from the day after acquisition, or the day of importation, until the day of discharge. For a newly colonized patient $j$, colonized on day $t$, a transmission source $s_j$ is chosen uniformly at random from the $C(t)$ positive patients present at the start of the day of colonisation. A set of genetic distances is generated according to the relationship between this patient and all previously observed patients with sequenced isolates. Under the importation structure model, distances are drawn from the Geom$(\gamma)$ or Geom$(\gamma_G)$ distributions, depending on whether the isolates are of the same type, or different type respectively. Under the transmission chain diversity model, distances are drawn from the Geom$(\gamma k^{\tau})$ or Geom$(\gamma_G)$  distributions, depending on whether the isolates belong to the same transmission chain ($\tau$ transmission events apart), or are unrelated, respectively.
At subsequent observation times resulting in positive results, genetic distances are generated accordingly. The first observation is assigned the same distances generated for the patient's importation/acquisition. Subsequent sequenced isolates differ from previous within-host sequences by $x$ SNPs, where $x\sim$Geom$(\gamma)$. 

\begin{figure}[!ht]
\centering
\includegraphics[width=8cm]{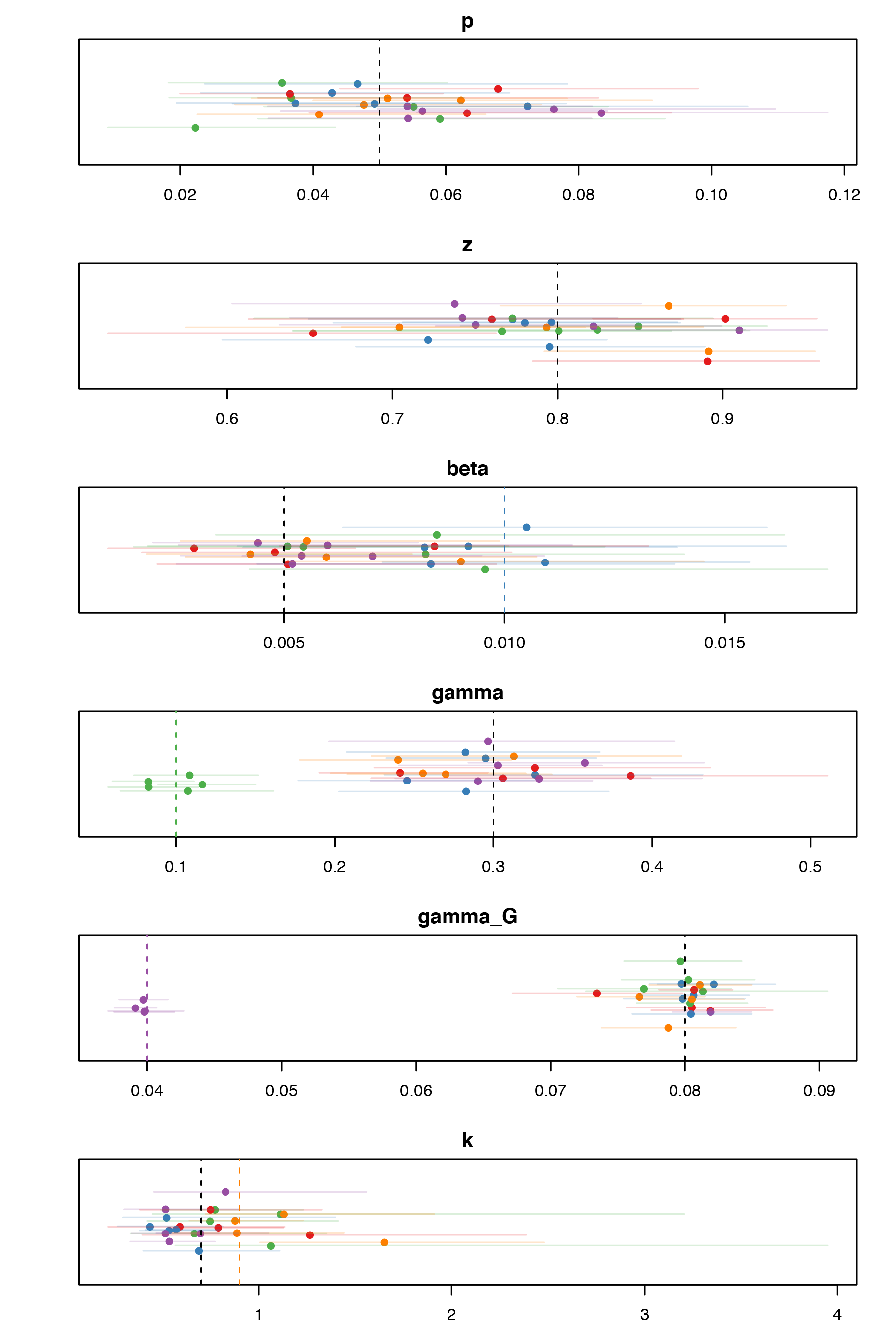}
\caption{Estimated parameters for the transmission chain diversity model for datasets simulated under various scenarios. Each panel shows posterior median estimates and 95\% credible intervals for each parameter. Colours represent the scenario under which the data were simulated; baseline (red), high transmission (blue), high within-host diversity (green), high between-chain diversity (purple), and high chain diversity parameter (orange). True values are shown as dashed vertical lines.
}
\label{fig:sim_recover2}
\end{figure}

For the simulations in this study, we used $D=7$, $x=3$, and simulated admissions over 250 days. In figures \ref{fig:sim_recover1} and \ref{fig:sim_recover2}, parameters estimated from simulated datasets are shown for the importation clustering model and the transmission diversity model respectively.

\subsection*{Simulation using `seedy'}

While we can recover parameters from data simulated under our models in many reasonable settings, we do not explicitly describe evolutionary dynamics, rather a process by which pairwise genetic distances are generated. While this allows additional flexibility in our framework, and avoids specification of processes which remain poorly understood, it is of interest to assess the performance of our analysis using data simulated under a more realistic evolutionary model. We used the R package `seedy' (Simulation of Evolutionary and Epidemiological Dynamics, \texttt{http://cran.r-project.org/web/packages/seedy/}) v1.2.1 to generate sequence data from a hospital outbreak. This software simulates pathogen evolution within- and between-host during a communicable disease outbreak, under a user-specified evolutionary model. We simulated trees as before, and passed infection times and routes to the \texttt{simfixoutbreak} function, which generated genome samples at designated times. We specified the effective population size to be 3000, transmission bottleneck size to be 1, and allowed the mutation rate and the importation diversity (that is, the expected genetic distance between two importations) to vary.

\begin{table}[h]
\tiny
\begin{center}
\begin{tabular}{l|cccccc|c}
Scenario&$p$&$z$&$\beta$&$\gamma$&$\gamma_G$&$k$&AUC\\
&&&&&&&($\Delta$ AUC)\\
\hline
$\mu=0.01,$&0.04&0.81&0.005&0.59&0.078&1.17&0.868\\
$d=5$&(0.02,0.06)&(0.65, 0.92)&(0.003, 0.009)&(0.47, 0.70)&(0.075, 0.081)&(0.79, 1.57)&(+0.168)\\
\hline
$\mu=0.03,$&0.07&0.71&0.010&0.49&0.041&0.78&0.98\\
$d=5$&(0.05,0.10)&(0.56, 0.81)&(0.006, 0.016)&(0.40, 0.60)&(0.039, 0.042)&(0.620, 1.02)&(+0.275)\\
\hline
$\mu=0.01,$&0.06&0.83&0.012&0.62&0.020&0.97&0.93\\
$d=25$&(0.04,0.09)&(0.72, 0.91)&(0.008, 0.017)&(0.51, 0.72)&(0.019, 0.021)&(0.80, 1.22)&(+0.224)\\
\hline
$\mu=0.03,$&0.08&0.81&0.011&0.25&0.018&0.66&0.938\\
$d=25$&(0.06, 0.11)&(0.72, 0.88)&(0.007, 0.015)&(0.21, 0.29)&(0.017, 0.018)&(0.54, 0.82)&(+0.179)\\
\hline
\end{tabular}
\caption{Posterior median estimates and 95\% credible intervals for parameters under data simulated under various scenarios. The variable $d$ is the expected number of SNPs between imported cases, while $\mu$ represents the per-generation mutation rate. Data were simulated under the parameters $p=0.05$, $\beta=0.008$ and $z=0.8$.
}
\label{tab:seedy_par}
\end{center}
\end{table}

\begin{figure}[!ht]
\centering
\includegraphics[width=12cm]{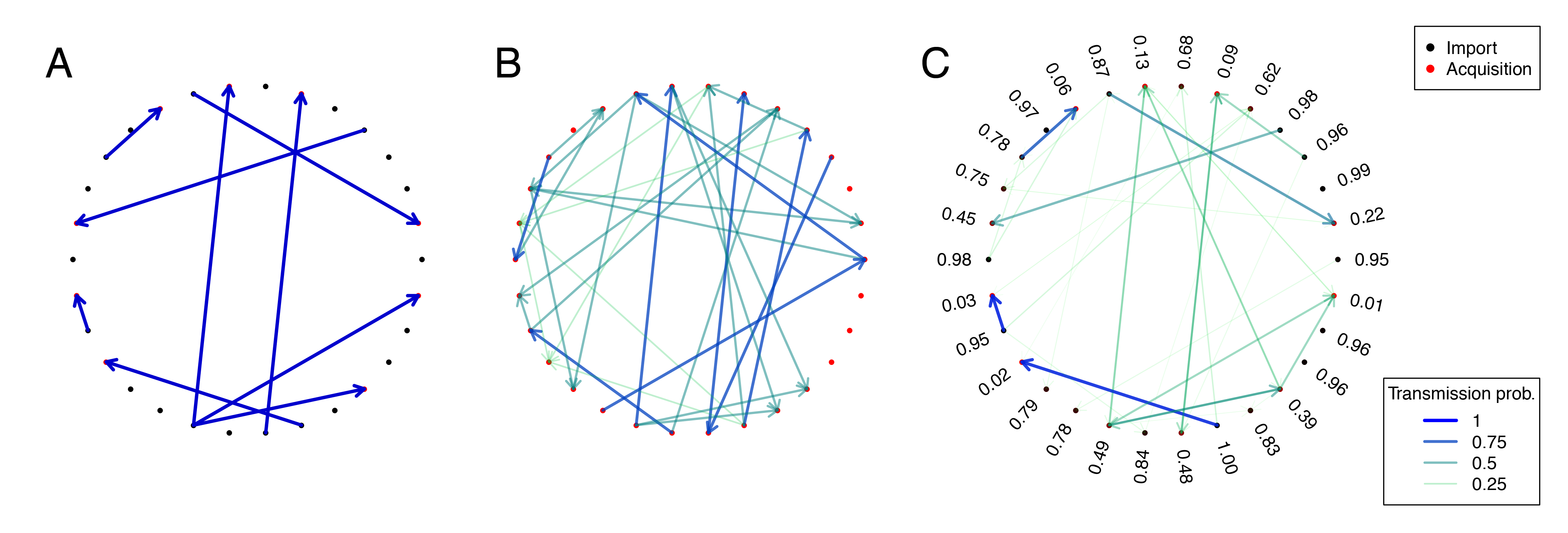}
\caption{Transmission tree recovery using genetic data simulated using the R package `seedy'. (A) The true transmission tree simulated as described in the simulations section ($p=0.05$, $\beta=0.008$). sampled genomes were simulated using the \texttt{simfixoutbreak} function in seedy (mutation rate 0.01 per generation, 20 pathogen generations per day, effective pathogen population size 3000, transmission bottleneck size 1). (B) The uninformed transmission tree, in which all possible transmission routes at time of infection are shown, and are weighted by the number of possible sources. (C) Inferred transmission tree, using the transmission diversity model.}
\label{fig:seedy}
\end{figure}

Figure \ref{fig:seedy} shows an example of a simulated outbreak, and the estimated routes of transmission, with genetic data simulated using seedy. The variable governing the expected distance between imported cases was a key factor in the performance of the model, with higher values typically leading to better network recovery. This effect is similar to lower values of $\gamma_G$ generating larger between-chain/group genetic distances under our models, and providing a greater discrimination between epidemiologically-linked hosts. Table \ref{tab:seedy_par} provides parameter estimates and $\Delta$ AUC values for simulated outbreaks.

\subsection*{Perturbation of the genetic distance matrix}

Errors in sequencing and alignment can lead to incorrectly observed genetic distances. In order to assess the impact of such errors, we perturbed genetic distance matrices generated from simulated outbreaks, and compared parameter estimates and inferred transmission routes. Matrices were perturbed by adding a Poisson-distributed `noise' variable to each pairwise distance; we considered expected errors of 0.1, 0.5, 1 and 2 SNPs.

\begin{figure}[!ht]
\centering
\includegraphics[width=8cm]{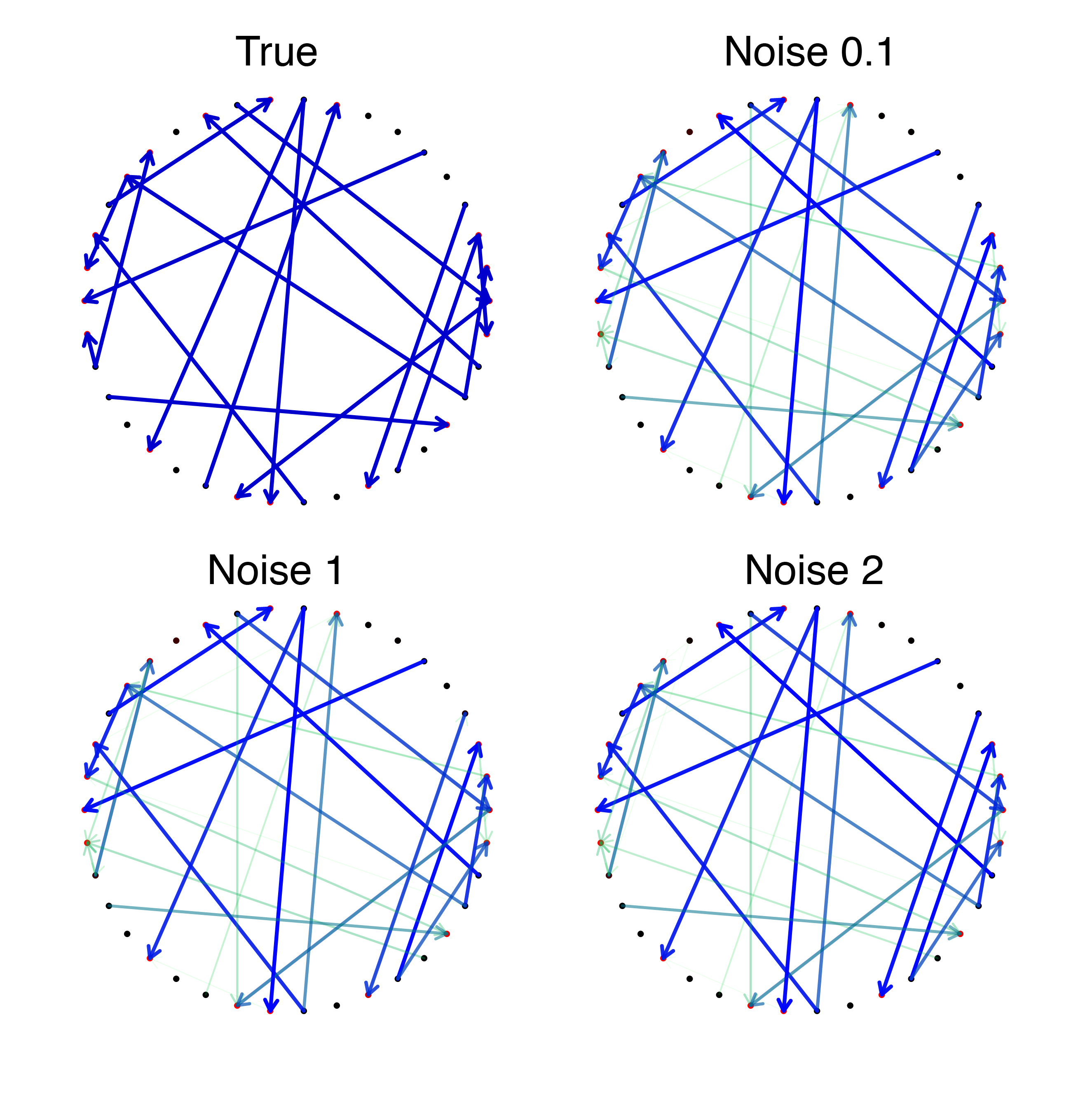}
\caption{A simulated transmission tree (top left), and recovered transmission routes under varying degrees of genetic distance observation error. The `noise' variable indicates the expected additional genetic distance observed on top of the true distance due to errors arising from sequencing and/or alignment.  The network was simulated under the transmission diversity model with $\gamma=0.3$, $\gamma_G=0.01$, $p=0.05$ and chain diversity parameter $k=0.7$.}
\label{fig:perturb_net}
\end{figure}

\begin{figure}[!ht]
\centering
\includegraphics[width=12cm]{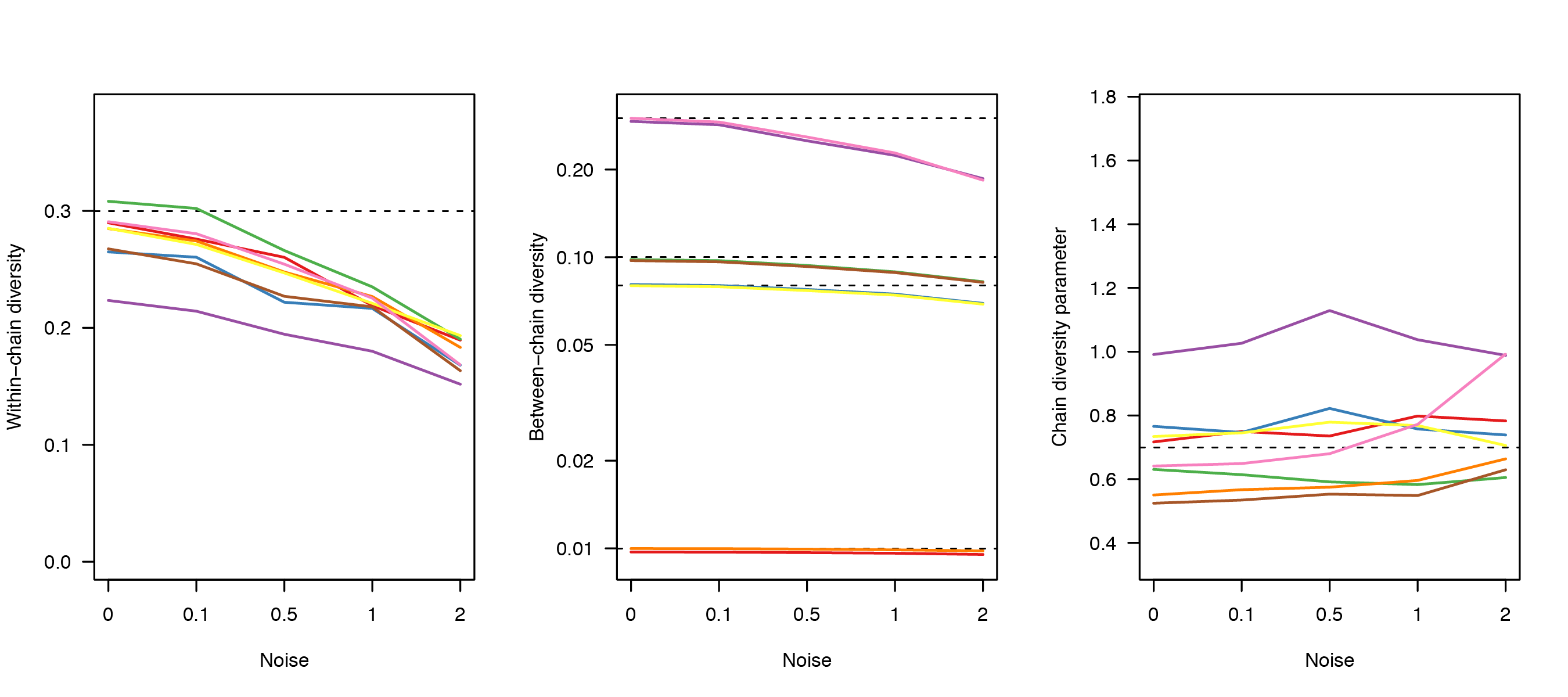}
\caption{Posterior median estimates for genetic diversity parameters under different levels of genetic distance error. Matrices were perturbed by adding a Poisson-distributed `noise' variable to each pairwise distance; we considered expected errors of 0.1, 0.5, 1 and 2 SNPs. Each coloured line corresponds to a different simulated scenario, with varying transmission rates ($\beta$=0.005, 0.008) and between-chain diversity values ($\gamma_G$=0.01, 0.08, 0.1, 0.3). True values of parameters are indicated with dashed horizontal lines; four different levels of $\gamma_G$ were used.}
\label{fig:perturb_par}
\end{figure}

Figure \ref{fig:perturb_net} shows an example simulated outbreak, and the transmission routes inferred using increasingly perturbed genetic distance matrix. In each of our simulation scenarios, we found that the transmission trees remained qualitatively similar, and compared to the tree generated under the unperturbed distance matrix, estimated trees had a $\Delta$ AUC between -0.01 and 0.01.

Figure \ref{fig:perturb_par} shows posterior parameter estimates for each simulation scenario under each level of matrix perturbation. While epidemiological parameters ($p$, $\beta$) remained unchanged by perturbation, the genetic diversity parameters fell in line with the error-induced increases in pairwise distances. The chain diversity parameter $k$ remained approximately the same, though with a wide credible interval.

\subsection*{Posterior predictive distributions}

We used posterior predictive distributions to assess model fit. Having fit our model to the Rosie hospital data, we repeatedly drew samples from the posterior distribution and used these to simulate new datasets, using the observed admission and discharge times which were not part of the modelling framework. We looked at the number of importations, judged as the number of patients whose first test result was positive, the number of acquisitions, defined to be the number of patients with a negative test followed by a positive test as well as the overall genetic diversity, the expected pairwise distance between any two isolates.

\begin{figure}[!ht]
\centering
\includegraphics[width=12cm]{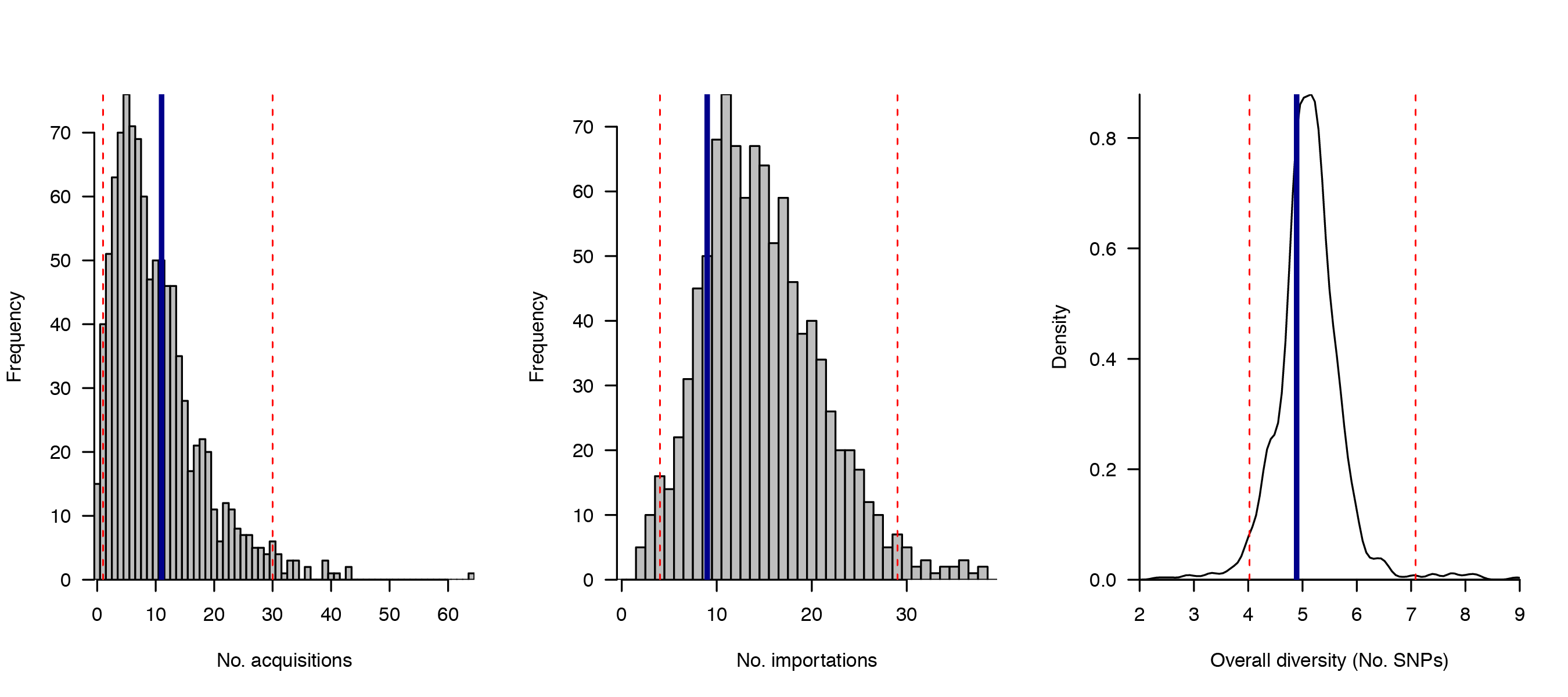}
\caption{Posterior predictive distributions for number of acquisitions (left) number of importations (centre) and the overall average pairwise diversity (right). Values observed from the Rosie hospital dataset are marked as vertical blue lines, while the bounds of the 95\% central quantile of the posterior predictive distribution are marked as dashed red lines.}
\label{fig:PPD}
\end{figure}

\begin{table}[h]
\begin{center}
\begin{tabular}{l|l}
Parameter&Estimate (95\% credible interval)\\
\hline
$p$&0.0109 (0.0057,0.0186)\\
$z$&0.765 (0.696,0.825)\\
$\beta$&0.0010 (0.0004,0.0019)\\
$\gamma$&0.206 (0.180,0.233)\\
$\gamma_G$&0.157 (0.143,0.171)\\
$k$&0.86 (0.56,0.99)\\
\end{tabular}
\caption{Posterior median estimates and 95\% credible intervals for the Rosie hospital dataset, with $k$ constrained to the interval $[0,1]$.
}
\label{tab:constrain}
\end{center}
\end{table}

\end{document}